\documentclass[reprint,superscriptaddress,amsmath,amssymb,aps,pre]{revtex4-1}

\usepackage{graphicx}
\usepackage{placeins}
\usepackage{float}
\graphicspath{plots}
\usepackage{dcolumn}
\usepackage{bm}
\usepackage{color}
\usepackage{bbold}
\usepackage{soul}
\usepackage{nicefrac}
\usepackage{flafter}     

\usepackage{catchfile}
\newcommand{\getenv}[2][]{%
\CatchFileEdef{\temp}{"|kpsewhich --var-value #2"}{\endlinechar=-1}%
\if\relax\detokenize{#1}\relax\temp\else\let#1\temp\fi}

\newcommand{\articlestoragepath}{/home/scratch/jstoeber/ArticleStorage/}
\getenv[\articlestoragepath]{ARTICLE_STORAGE}
\IfFileExists{pdf_article_link.sty}{\usepackage{pdf_article_link}}

\newcommand{\HIDDEN}[1]{}

\newcommand{\R}{\mathbb{R}}
\newcommand{\Z}{\mathbb{Z}}
\newcommand{\N}{\mathbb{N}}
\newcommand{\C}{\mathbb{C}}

\providecommand*{\oneD}{\textsc{1d}}
\providecommand*{\twoD}{\textsc{2d}}

\providecommand*{\threeD}{\textsc{3d}}
\providecommand*{\threeDD}{\textsc{3D}}
\providecommand*{\fourD}{\textsc{4d}}
\providecommand*{\fourDD}{\textsc{4D}}

\newcommand{\dtori}{\ensuremath{d_{\text{tori}}}}
\newcommand{\onetori}{\oneD{} tori}

\newcommand{\M}{\ensuremath{\mathcal{M}}}
\newcommand{\D}{\ensuremath{\mathrm{D}}}
\newcommand{\mv}[1]{\ensuremath{\bm{#1}}}
\newcommand{\tr}[1]{\ensuremath{\text{tr}\,{#1}}}

\newcommand{\pa}{(A)}
\newcommand{\pb}{(B)}
\newcommand{\pc}{(C)}
\newcommand{\pd}{(D)}
\newcommand{\pe}{(E)}
\newcommand{\pf}{(F)}

\newcommand{\eqnref}[1]{Eq.~(\ref{#1})}
\newcommand{\figref}[1]{Fig.~\ref{#1}}
\newcommand{\Figref}[1]{Figure~\ref{#1}}

\newcommand{\secref}[1]{section~\ref{#1}}
\newcommand{\Secref}[1]{Section~\ref{#1}}

\usepackage{color}

\newcommand{\ABJS}[1]{AB$\to$JS}
\newcommand{\JSAB}[1]{JS$\to$AB}

\newcommand{\movierefall}{For a rotating view see
\href{http://www.comp-phys.tu-dresden.de/supp/}{
http://www.comp-phys.tu-dresden.de/supp/}.}

\let\OLDthebibliography\thebibliography
\renewcommand\thebibliography[1]{
  \OLDthebibliography{#1}
  \setlength{\parskip}{0pt}
  \setlength{\itemsep}{0pt plus 0.3ex}
}

\makeatletter
\let\Hy@backout\@gobble
\makeatother

\begin{document}

\title{Geometry of complex instability and escape
       in four-dimensional symplectic maps}

\author{Jonas St\"ober}
\affiliation{Technische Universit\"at Dresden, Institut f\"ur Theoretische
             Physik and Center for Dynamics, 01062 Dresden, Germany}
\affiliation{Max-Planck-Institut f\"ur Physik komplexer Systeme, N\"othnitzer
Stra\ss{}e 38, 01187 Dresden, Germany}

\author{Arnd B\"acker}
\affiliation{Technische Universit\"at Dresden, Institut f\"ur Theoretische
             Physik and Center for Dynamics, 01062 Dresden, Germany}
\affiliation{Max-Planck-Institut f\"ur Physik komplexer Systeme, N\"othnitzer
Stra\ss{}e 38, 01187 Dresden, Germany}

\date{\today}

\begin{abstract}
In four-dimensional symplectic maps complex instability of
periodic orbits is possible, which cannot occur in the two-dimensional case.
We investigate the transition from stable to complex unstable dynamics of a
fixed point under parameter variation. The change in the geometry of regular
structures is visualized using \threeD\ phase-space slices
and in frequency space using the example of two coupled standard maps.
The chaotic dynamics is studied using escape
time plots and by computations of the \twoD\ invariant manifolds associated
with the complex unstable fixed point. Based on a normal-form description, we
investigate the underlying transport mechanism by visualizing the escape paths
and the long-time confinement in the surrounding of the complex unstable
fixed point. We find that the escape is governed by the transport along the
unstable manifold across invariant planes of the normal-form.
\end{abstract}

\maketitle

\section{Introduction}
\label{sec:introduction}

There are different ways in which orbits of a dynamical system
may become unstable under variation of some parameter.
One famous example is the Hamiltonian-Hopf bifurcation
as has first been studied for the
triangular equilibrium points of the planar circular
restricted three-body problem \cite{Bro1969,Had1975},
for which instability occurs beyond a critical mass ratio \cite{Sic2010}.
This is also found for many other examples in celestial and galactic dynamics
\cite{Mag1982,MarPfe1987,Heg1985,ConMag1985,PatZac1990,OllPacVil2004,
  Hanvan2005, KatPatCon2011,PatKat2014b}
for the hydrogen atom  \cite{EfsCusSad2004, LahRoy2001, OllPac2018},
in the context of molecular dynamics
\cite{FarFou1990,DiaEgeFervanVer2010},
and is also of relevance to particle accelerators \cite{HowLicLieCoh1986}.
The impact of the Hamiltonian-Hopf bifurcation on the phase space geometry
has been studied in much detail in
Refs.~\cite{Mee1985, Cra1991, PapConPol1995}.
Additional insight is provided
by computations of invariant manifolds and normal-form descriptions
\cite{McSMey2003, OllPacVil2004b, OllPacVil2005b, FonSimVie2019}.
For further results see e.g.\ Refs.~\cite{Wen2005, BroHanHoo2007, OllPacVil2008, VitBroSim2011}. The impact in quantum mechanical systems has been
investigated in Ref.~\cite{ConFarPapPol1994}.

Often its is helpful to reduce the time-continuous dynamics
to a discrete-time mapping by means of a Poincar\'e section.
For conservative Hamiltonian systems with three degrees of freedom
this leads to the study of four-dimensional (\fourD) symplectic maps,
which are therefore of importance of many areas of physics.
Similar to the Hamiltonian case, a transition
from stable to complex unstable dynamics is possible
for \fourD{} (and higher-dimensional) symplectic maps
\cite{HowMac1987,Sko2001b}.
This has been investigated in detail in the pioneering work
\cite{Pfe1985a, Pfe1985b} for a variant of the \fourD{} coupled
standard map \cite{Fro1971}.
In such a transition to complex unstable dynamics
two eigenvalue pairs of the linearized dynamics
collide on the unit circle and afterwards form a so-called Krein quartet.
This may only happen, if the Krein signature is mixed \cite{HowMac1987}.
A distinctive feature is the spiraling motion in the surrounding
of a complex unstable periodic point
\cite{Heg1985,ConFarPapPol1994}.
Moreover, it was found that commonly
an extended region around a complex unstable fixed point emerges
to which the dynamics is confined for rather long times
\cite{JorOll2004, KatPatCon2011, ZacKatPat2013,  DelCon2016}.
Important approaches to understand the complex unstable dynamics
are based on computations of the invariant manifolds
\cite{RoyLah1991, JorOll2004, DelCon2016}
and normal form descriptions \cite{BriCusMac1995, FonSimVie2015, OllPac2018}.
Hamiltonian-Hopf bifurcations have also been studied in much
detail for reversible maps,
see e.g.\ Refs.~\cite{LahBhoRoy1998,BhoRoyLah1993b}.

In this paper, we investigate how the transition from stability to
complex instability of a fixed point affects the geometry of invariant
objects in its surrounding in the phase space of a \fourD{} symplectic map.
This transition is accompanied by the possibility that orbits can escape
from the vicinity of the fixed point
which is quantified by the average escape times
of an ensemble of orbits.
The underlying escape mechanism is investigated in terms of the
geometry of the stable and unstable manifolds.
We provide evidence that the escape occurs across
the invariant planes of the normal-form description
showing that it is a genuinely higher-dimensional mechanism.

The text is organized as follows.
In Sec.~\ref{sec:complex_unstable_dynamics} we
recall some fundamental properties of linear stability of fixed points and the
requirements for complex instability in \fourD{} symplectic maps.
Section~\ref{subsec:normal_form} summarizes a normal-form description
for the transition to complex instability as introduced
in Ref.~\cite{BriCusMac1995}.
In Sec.~\ref{sec:transition_cu} we introduce a variant
of the four-dimensional coupled standard map and define a set of
parameters for investigating the transition from elliptic-elliptic stability to
complex instability. We visualize the dynamics in the \fourD{} phase space
using \threeD\ phase-space slices \cite{RicLanBaeKet2014}
which is complemented by a frequency space representation
\cite{Las1990, Las1993, Las1993b}.
The escape dynamics is investigated in Sec.~\ref{sec:escape}
for an ensemble of initial conditions close to the complex unstable
fixed point.
To explain the underlying mechanism we
compute the stable and unstable manifolds associated with the complex unstable
fixed point by utilizing the parametrization method \cite{CabFonLla2003a,
  CabFonLla2003b, CabFonLla2005}.
The dynamics of the ensemble suggests that the escape occurs
across invariant planes of the corresponding normal-form description.
\Secref{sec:summary} gives a summary and outlook.

\section{Complex unstable dynamics}
\label{sec:complex_unstable_dynamics}

\subsection{Linear stability in \fourDD\ maps}
\label{subsec:linear_stability}

In this section we collect some important results on the stability
of fixed points in \fourD\ symplectic maps \cite{HowMac1987},
the Krein collision \cite{Mos1958,ArnAve1968,Kre1983,HowMac1987}
and its normal-form description \cite{BriCus1993, BriFur1993, BriCusMac1995}.
A map $\M: \R^{4} \to \R^{4}$ is called symplectic
if its Jacobian matrix $\D\M$ fulfills $\D\M^T J \D\M = J$,
where $J=\begin{pmatrix} 0 & -I \\ I & 0 \end{pmatrix}$ is the
$4\times 4$ Poisson matrix with $I$ being the $2\times2$ identity matrix.
An immediate consequence is that a symplectic map
is volume preserving as $\det(\D\M) = 1$.
The dynamics in the vicinity of a
fixed point, i.e., a point $\mv{z}^{*}$ that satisfies $\M\mv{z}^{*} =
\mv{z}^{*}$, is given by the linearized map $\D\M$.
The symplecticity of \M\ implies that the characteristic polynomial
$P(\lambda)$ of $\D\M$ is reflexive so that coefficients
of $P$ come in palindromic form.
For a \fourD{} symplectic map this can be written as
\begin{equation}
 P(\lambda) = \lambda^4 - A\lambda^3 + B\lambda^2 -A\lambda +1,
\label{eq:char_polynomial}
\end{equation}
where $A=\tr{(\D\M)}$ and $2\,B=A^2-\tr{(\D\M^2)}$. As consequence, the
eigenvalues $\lambda_j$ with  $j\in\{1,2\}$ are restricted to
either hyperbolic pairs $\lambda_j, \lambda_j^{-1} \in \R$,
elliptic pairs of $\lambda_j, \bar{\lambda}_j \in \C$ with $|\lambda_j| = 1$ or
a Krein quadruplet of complex eigenvalues
$\lambda, \lambda^{-1}, \bar{\lambda}, \bar{\lambda}^{-1} \in \C$ with
$|\lambda| \neq 1$.

This gives a total of four possible stability types,
namely elliptic-elliptic (EE), elliptic-hyperbolic (EH),
hyperbolic-hyperbolic (HH) and complex instability (CU).
These stability types can be
distinguished by introducing the stability index of an eigenvalue pair
$\rho=\lambda_j+\lambda_j^{-1}$ and reducing the characteristic polynomial in
\eqnref{eq:char_polynomial} to
\begin{equation}
 R(\rho)=P(\lambda)\,\lambda^{-2}=\rho^2-A\rho+B-2.
 \label{eq:reduced_polynomial}
\end{equation}
As shown in Ref.~\cite{HowMac1987}, different regimes of
stability follow from \eqnref{eq:reduced_polynomial} in dependence
on $A$ and $B$. The linearized map $\D\M$ is spectrally stable
if and only if all roots of $R(\rho)$ are real
and within the interval $[-2, 2]$.
Therefore $R(\pm2)=0$ yields two stability boundaries, namely
\begin{equation}
 B=\pm2A-2.
 \label{eq:linear_boundaries}
\end{equation}
Crossing either of these boundaries corresponds to a saddle-center (SC) or
a period-doubling (PD) bifurcation, respectively. Another boundary corresponds
to the roots of $R(\rho)$ becoming complex, which occurs when the
discriminant of the reduced characteristic polynomial $\Delta \, R(\rho) =
(\rho_1 - \rho_2)^2=0$. This gives the so-called Krein parabola (KP)
\begin{equation}
 B=A^2/4+2.
 \label{eq:krein_parabola}
\end{equation}

\begin{figure}
 \includegraphics{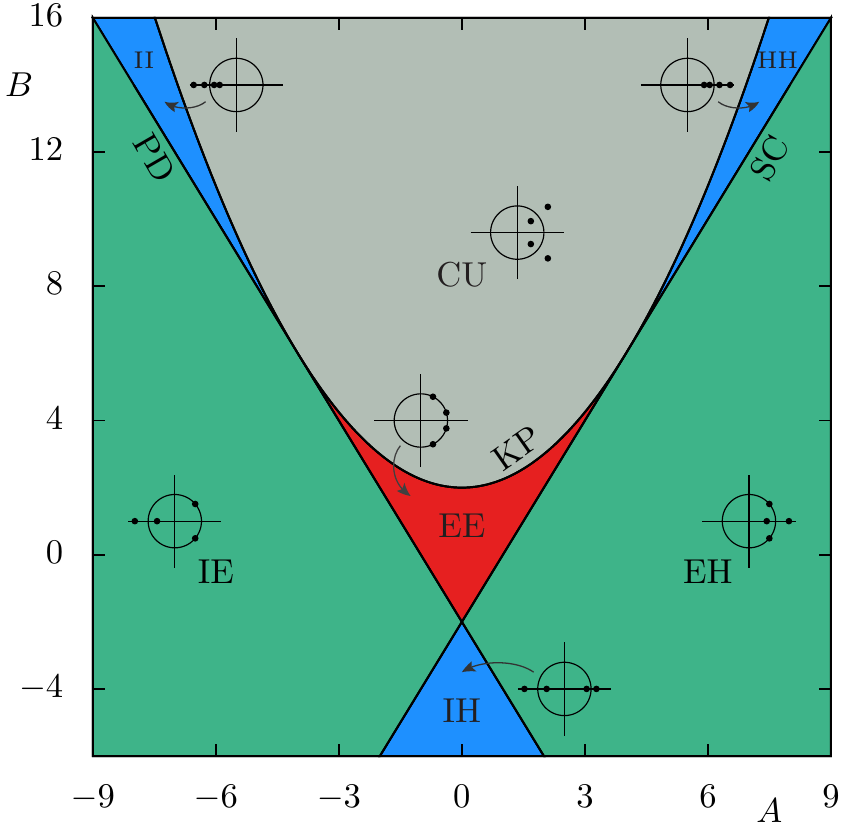}
 \caption{Stability of a fixed point in dependence on the coefficients $A$ and
   $B$ of the characteristic polynomial \eqref{eq:char_polynomial}
   of the linearized map $\D\M$. The regions
   correspond to combinations of  elliptic (E), hyperbolic (H), and
   inverse hyperbolic (I), or complex unstable (CU).
   The regions are seperated by the period-doubling line
   (PD), saddle-center line (SC), and the Krein  parabola (KP).}
 \label{fig:broucke}
\end{figure}

The possible stability types for an arbitrary fixed point
of a \fourD\ map in dependence of $A$ and $B$ can be displayed
in the so-called Broucke diagram \cite{Bro1969, HowMac1987},
see \figref{fig:broucke}.
The three stability boundaries SC, PD, and KP lead to
seven stability regions corresponding to
complex instability (CU) and the different
combinations of the elliptic (E), the hyperbolic (H) case,
and the inverse hyperbolic (I) case, for which the
eigenvalue pair lies on the negative real axis.
The corresponding arrangement of the eigenvalues of
the linearized map are shown as small insets.

For an EE fixed point the surrounding
consists of a two-parameter (Cantor) family of
\twoD{} tori as expected from Kolmogorov-Arnold-Moser (KAM) theory.
The \twoD{} tori are organized around
one-parameter (Cantor) families of elliptic \onetori{}.
These families are commonly referred to
as Lyapunov families, based on the analogy
to the Lyapunov center theorem for Hamiltonian flows \cite{MeyOff2017}.
Such families of \onetori{} have been studied in detail,
see e.g.\ Refs.~\cite{Gra1974, Zeh1976, JorVil1997, JorVil2001, JorOll2004,
  LanRicOnkBaeKet2014,OnkLanKetBae2016}.
As the families of elliptic \onetori{} form the `skeleton'
of the surrounding regular dynamics, they
allow for a convenient way to understand the change
in geometry occurring when an EE fixed point becomes CU,
as will be illustrated below in Sec.~\ref{subsec:phase_space_slices}.

\subsection{Krein collision}
\label{subsec:krein_collision}

\begin{figure}
 \includegraphics[width=8.6cm]{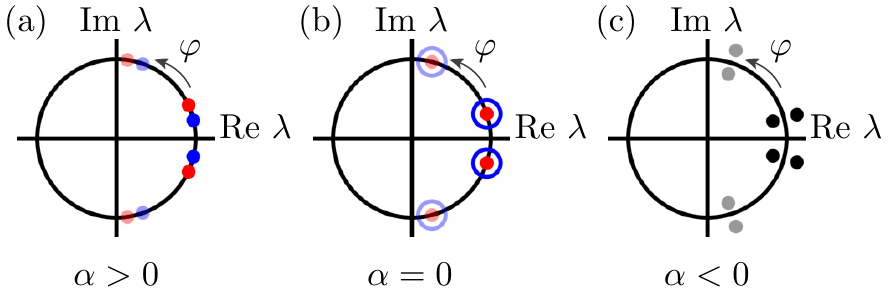}
 \caption{Krein collision of two elliptic eigenvalue pairs (red and blue
circles) in dependence of $\alpha$. The eigenvalues coalesce for $\alpha=0$
and split off the complex unit circle for $\alpha<0$ forming a Krein
quadruplet (black circles). For non-zero angle $\varphi$ the
location of the Krein collision is move along the unit circle.}
 \label{fig:krein_collision}
\end{figure}

As seen from Broucke's diagram in \figref{fig:broucke}, there are only three
possible ways to enter the CU regime, namely the transition from a) the
elliptic-elliptic (EE), or b) the hyperbolic-hyperbolic (HH or II) stability
regions through the Krein parabola, or c) through the intersection points of
the Krein parabola with either the saddle-center or the period-doubling
boundary at $(A, B)=(\pm4, 6)$.
The most interesting case is the transition of
an elliptic-elliptic fixed point, i.e.\ case a),
as  illustrated in Fig.~\ref{fig:krein_collision}
in dependence on some parameter $\alpha$,
which controls the transition. For $\alpha>0$, two elliptic
eigenvalue pairs approach each other on the complex unit circle until they
coalesce at $\alpha=0$. For $\alpha<0$, the eigenvalues split off the unit
circle and form a Krein quadruplet.

Whether the eigenvalue pairs of an EE fixed point for a given map can leave
the unit circle or pass through each other while staying on the unit circle
depends on the so-called Krein signature.
This is given by the signature $(m_+, m_-)$ of the quadratic form
\begin{equation} \label{eq:quadratic-form}
   q(x) = x^T\,J\,\D\M\,x,
\end{equation}
which can for example be computed numerically
from the eigenvalues of the symmetric matrix
$\frac{1}{2}\left(J\,\D\M + (J\,\D\M)^T\right)$,
where $m_+$ is the number of positive and
$m_-$ is the number of negative eigenvalues.
If $m_+=0$ or $m_-=0$
then the fixed point cannot loose stability and
stays elliptic-elliptic. Conversely, the fixed point may loose its stability
and become complex unstable if the signature is mixed.
Note that the quadratic form \eqnref{eq:quadratic-form} allows the
construction of an invariant of the linearized dynamics as
\begin{equation}\label{eq:quadratic_invariant}
 q(x) = x^T\,J\,\D\M\,x = (\D\M \, x)^T\, J\, \D\M \, (\D\M \, x)
\end{equation}
is preserved under $\D\M$ \cite{Pfe1985a}.

The geometric interpretation of the Krein signature
becomes more clear when considering
the signature of a multiplier $\lambda$ on the unit circle,
\begin{equation}
  \sigma(\lambda) = \text{sgn}\, q(u),
\end{equation}
where $u$ is any real vector in the eigenspace of $\lambda$.
If eigenvalues with the same signature collide on the unit circle,
they cannot split off to form a  Krein quartet.
Explicitly, consider a \fourD{} symplectic map which is uncoupled,
i.e.\ $\M (p_1, p_2, q_1, q_2) = (p_1', p_2', q_1', q_2')$
with $(p_1', q_1') = \M_1 (p_1, q_1)$ and
$(p_2', q_2') = \M_2 (p_2, q_2)$.
Then using the quadratic form \eqref{eq:quadratic-form}
and $(1,0,0,0)$ and $(0, 1, 0, 0)$ as vectors of the corresponding
eigenspaces the signatures are given by
$\sigma(\lambda_i) =  \text{sgn}( (\D\M_i)_{12} )$.
Therefore the fixed point can only become complex unstable
under some generic coupling if \cite{Con1986,ConGio1988}
\begin{equation} \label{eq:Krein-DP-product}
  \text{sgn}((\D\M_1)_{12})  \;  \text{sgn}((\D\M_2)_{12}) < 0 .
\end{equation}
This reflects the counter-rotating nature of the dynamics
in the two independent subspaces,
similar to the Cherry-Hamiltonian
describing two counter-rotating harmonic oscillators \cite{Che1928}.

Furthermore, a mixed Krein signature implies that the linearized map of the
coalesced eigenvalues has non-trivial Jordan blocks of the shape $m_+\times\,
m_+$ and $m_-\times\,m_-$ while the matrix can be diagonalized if the
signature is positive or negative definite. Thus, the linearization takes
either the form \cite{Pfe1985a}
\begin{equation}
 \begin{pmatrix}
  \lambda & 1 & 0 & 0\\ 0 & \lambda & 0 & 0\\
  0 & 0 & \bar{\lambda} & 1\\ 0 & 0 & 0 & \bar{\lambda}
 \end{pmatrix}\quad\text{or}\quad
 \begin{pmatrix}
  \lambda & 0 & 0 & 0\\ 0 & \lambda & 0 & 0\\
  0 & 0 & \bar{\lambda} & 0\\ 0 & 0 & 0 & \bar{\lambda}
 \end{pmatrix}
\end{equation}
where $\lambda=e^{\textrm{i}\theta}$ and $\theta \in \, ]0, \pi[$. Beside this,
in case b) the signature is always mixed. Thus for an II or HH fixed point
there is no constraint to enter the CU region.

\subsection{Normal form description}
\label{subsec:normal_form}

To understand the geometry of regular and invariant
structures around a CU fixed point, it is helpful to consider
a non-linear normal-form description \cite{BriCusMac1995},
of which we now summarize the main aspects. Consider a symplectic map \M,
\begin{equation}
 \mv{x}'=\M(\mv{x}, \alpha, \varphi),
\end{equation}
with $\mv{x}, \mv{x}' \in \R^4$ and parameters $\alpha, \varphi \in \R$.
The fixed point is assumed to be at the origin
$\mv{x}=\mv{0}$ such that $\M(\mv{0}; \alpha, \varphi)=0$
for arbitrary $\alpha$ and $\varphi$.
Furthermore, the eigenvalues
of the linearized map $\D\M(\mv{0}; 0, 0)$ are assumed to coalesce
ot $\lambda = \exp{(\pm\textrm{i}\theta)}$ with $\theta=2\pi\nu$ and irrational
$\nu \in ]0, \nicefrac{1}{2}[$.
Note that the case of the rational Krein collision is for example
considered in Ref.~\cite{BriFur1993}. The collision is
controlled by the parameters $\alpha$ and $\varphi$ as shown in
\figref{fig:krein_collision}.
The parameter  $\alpha$ controls the transition from the
elliptic-elliptic eigenvalue pair for $\alpha>0$ to the complex unstable
quadruplet for $\alpha<0$.
The angle $\varphi$ rotates the angle of the Krein collision on the complex
unit circle.

In case of the irrational Krein collision with
$\alpha=0$ and $\varphi=0$, the linearized map has non-trivial Jordan blocks
and can be brought into a Williamson normal-form $L_0$ by a
symplectic transformation $T$
\begin{align}
 &T^{-1}\,\D\M\;T = L_0(\mv{0}; 0, 0) =
 \begin{pmatrix} R_\theta & \epsilon R_\theta\\ 0 & R_\theta
 \end{pmatrix}
\end{align}
where $\epsilon=\pm1$ and
\begin{align}
R_\theta =
 \begin{pmatrix} \cos{\theta} & \sin{\theta}\\ -\sin{\theta} & \cos{\theta}
 \end{pmatrix}.
\end{align}
For $\alpha\neq0$ and $\varphi \neq0$, the Williamson normal-form has a
transversal two-parameter unfolding, i.e., there is a two-parameter family of
matrices that preserve the symplectic form and describes the transition from
stability to complex instability via the Krein collision given by
\cite{BriFur1993}
\begin{equation}
 L=L_0(\mv{0},\alpha, \varphi)=
 \begin{pmatrix}
  (1-\epsilon\alpha)R_{\theta+\varphi}&\epsilon R_{\theta+\varphi}\\
  -\alpha R_{\theta+\varphi}&R_{\theta+\varphi}
 \end{pmatrix}.
\end{equation}
With that, the transformed map $\widetilde{\M}$ in the new coordinates
$\mv{y}$ can be represented as a formal power series
\begin{equation}
 \mv{y}'=\widetilde{\M}(\mv{y},\alpha,\varphi)\approx L +
\Phi_2(\mv{y},\alpha,\varphi) + \ldots
 \label{eq:power_series}
\end{equation}
where $\Phi_j(\mv{y},\alpha,\varphi)$ are vector-valued polynomials of degree
$j$. In Refs.~\cite{BriCus1993, BriFur1993, BriCusMac1995} it is shown that
\eqnref{eq:power_series} can be normalized by utilizing a symplectic
diffeomorphism $\Psi_j(\mv{y})$ such that
$\Psi_j^{-1}\circ\widetilde{\M}\circ\Psi_j$ is in normal-form with respect to
$L$ up to order $j$ for arbitrary $j\in\N$.

As a result, one gets
the non-linear normal-form
\cite{BriCusMac1995}
\begin{equation}
 \begin{pmatrix}x'\\y'\end{pmatrix} =
 \begin{bmatrix}(1-\epsilon h)R_{\theta+\nu}&\epsilon R_{\theta+\nu}\\
                -h R_{\theta+\nu}&R_{\theta+\nu}
 \end{bmatrix}\begin{pmatrix}x\\y\end{pmatrix}
 \label{eq:poisson_map}
\end{equation}
with $(x, y)=(x_1, x_2, y_1, y_2) \in \R^4$. The parameters
\begin{align*}
 h & = \alpha + b_1 X + b_2I + \ldots\\
 \nu & = \varphi + b_2 X + b_3 I + \ldots.
\end{align*}
are derivatives of a deduced Hamiltonian generating function with respect to
the coordinates $X=x_1^2+x_2^2$ and $I=y_1x_2-x_1y_2$, respectively.
For our purposes we truncate the series of the generating function after
quadratic order and obtain $\widetilde{\nu}=\varphi$ and
$\widetilde{h}\approx\alpha+bX$ where $\widetilde{h}$ is scaled with respect
to $X$ such that $b=\pm1$.

The normal-form \eqnref{eq:poisson_map}
is guaranteed to be equivariant to a symmetry operation
\cite[Thm.~2.7]{BriCus1993}, i.e.\ the normal-form commutes with the action of
a symmetry group. A straightforward computation reveals that
\eqnref{eq:poisson_map} is $S^1$-equivariant where the symmetry transformation
acts as rotation on all coordinates $(R_\gamma x, R_\gamma y)$ for $\gamma \in
[0, 2\pi[$. The corresponding invariant of \eqnref{eq:poisson_map} is
$I(x', y')=I(x, y)= y_1x_2-x_1y_2$. Consequently, the \fourD\ nonlinear normal
form map can be reduced further by introducing new coordinates.

\begin{figure}
 \includegraphics{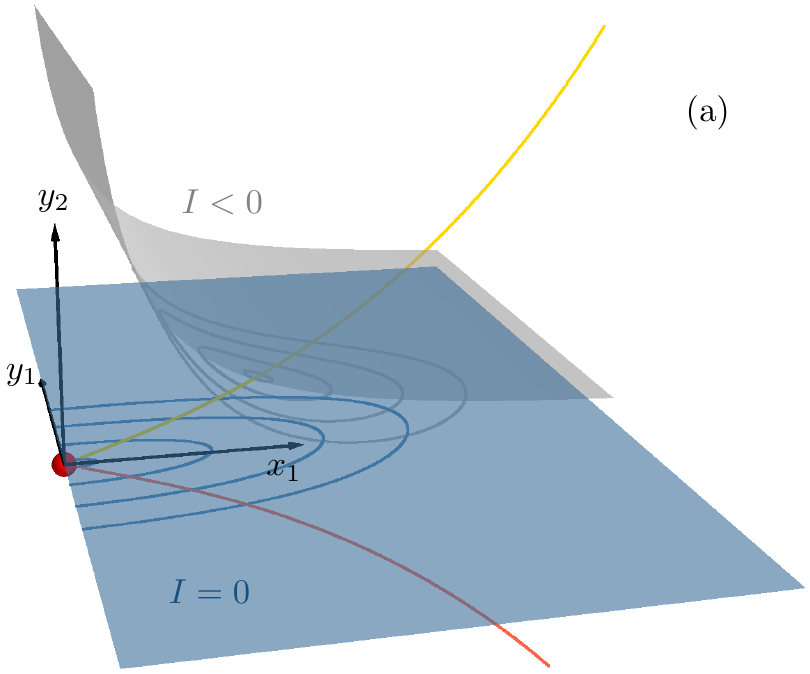}
 \includegraphics{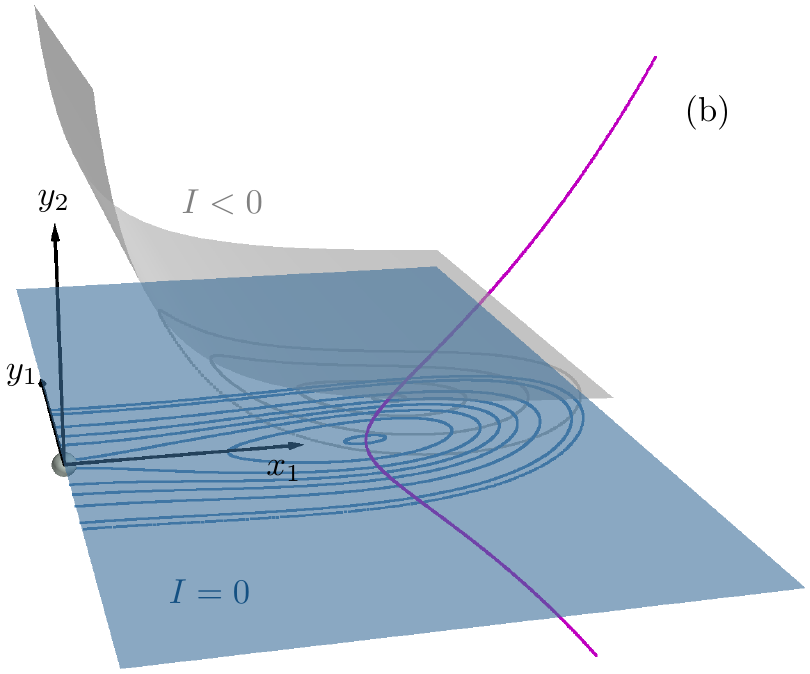}
 \caption{The reduced Poisson map from \eqnref{eq:poisson_map} in $(x_1,
y_1,y_2)$ coordinates. The sphere in the origin denotes the trivial
fixed point while the gray and the blue planes visualize the $I=-0.015$ and
the $I=0$ plane, respectively. The shown orbits correspond to the same plane
as their color indicates. The non-trivial periodic points of the reduced map
are depicted as orange and yellow dots for the EE case (a) for $\alpha>0$ and
as magenta dots for the CU case (b) for $\alpha<0$.
}
 \label{fig:poisson_map}
\end{figure}

Hence, we take advantage of the symmetry and visualize the dynamics of
\eqnref{eq:poisson_map} in the hyperplane $x_2=0$,
see Fig.~\ref{fig:poisson_map}.
Note that the full
dynamics can be re-obtained by applying the symmetry operation, i.e.\ by
simultaneous rotation in the $x$ and $y$ coordinates,
see Ref.~\cite[Eq.~(3.1)]{BriCusMac1995}.
For the sake of clarity, we stick to the half-space with $x_1\ge0$ since the
other half can be obtained by the transformation $(x_1,y_1)\mapsto(-x_1,y_1)$.
Furthermore, without loss of generality we fix the parameters $\epsilon=1$ and
$b=1$. Firstly, we consider the case $I=0$, i.e.\ $0 = I = -x_1y_2$.
Without loss of generality, we choose $y_2=0$ and
\eqnref{eq:poisson_map} reduces to a \twoD\ map $f(x_1, y_1)\mapsto(x_1',y_1')$
\begin{subequations}
\begin{align}
 x_1'&=|(gx_1+y_1|\\y_1'&=(\widetilde{h}x_1-y_1)\,\text{sign}(gx_1+y_1)
\end{align}
\end{subequations}
with $g=1-\epsilon h$. This map has two periodic points, namely a trivial
fixed point at $(0, 0)$ which is the original fixed point of \M\ and
for $\alpha<0$ a non-trivial period-two periodic point
at $(\sqrt{-\alpha/b}, 0)$. A stability
investigation reveals that the trivial fixed point becomes
unstable for negative $\alpha$ as expected.
In contrast, the non-trivial periodic point only
exists when $\alpha\le0$ and is always stable. This particular situation in
the $I=0$ plane corresponds to the typical behavior of a period-doubling
bifurcation in a \twoD{} symplectic map, for which  a periodic point looses its
stability and  a stable periodic point of the twice the period is created,
see e.g.\ \cite{Mey1970, GreMacVivFei1981, Kay1993, MeyHalOff2009}.

For the second case $I\neq0$, the coordinate $y_2$ is given by the invariant
$I$. Thus, \eqnref{eq:poisson_map} reduces to a \twoD\ map with all structures
living on a hypercolic cylinder $y_2=-I/x_1$ in the reduced phase space. The
map takes the form
\begin{subequations}
\begin{align}
 x_1'&=\sqrt{(gx_1+y_1)^2+\nicefrac{I^2}{x_1^2}}\\
 y_1'&=\frac{(gx_1+y_1)(y_1-\widetilde{h}x_1)+\nicefrac{I^2}{x_1^2}}{x_1'}.
\end{align}
\end{subequations}
In this case, there is only one non-trivial period-two point,
which is given by an implicit equation that we solve numerically.

\Figref{fig:poisson_map}(a) shows the reduced phase space in $(x_1, y_1,
y_2)$ coordinates for $\alpha>0$, i.e., the stable case. The red sphere
represents the trivial fixed point which is elliptic-elliptic in this case. The
blue and the gray planes as well as the orbits in the same color correspond to
$I=0$ and $I=-0.015$, respectively. As long as $\alpha$ is positive, there
exits only one fixed point in the $I=0$ plane. For
$I>0$ and $I<0$ we get a continuous family of non-trivial periodic
points, shown as red and yellow curves, respectively,
which are both attached to the trivial fixed point at the origin.

\Figref{fig:poisson_map}(b) shows the reduced phase space for $\alpha<0$. The
trivial fixed point (gray sphere) has become unstable and the family of
non-trivial periodic points of the $I\neq0$ plane are detached from the origin
similar to a period-doubling bifurcation in a \twoD\ map. In this way, this
family with its surrounding stable \onetori{} forms a foliated tube-like object
in
phase space.

\section{Transition to complex instability}
\label{sec:transition_cu}

\subsection{\fourDD\ map with CU fixed point}
\label{subsec:4d_map}

The usual \fourD{} standard map \cite{Fro1970a, Fro1972},
which has been investigated in much detail,
see e.g.\ Refs.~\cite{Har1999, GuzLegFro2002, CelFalLoc2004, ZacKatPat2013,
RicLanBaeKet2014},
does not allow for CU fixed points.
A modified \fourD{} standard map has been introduced in Ref.~\cite{Pfe1985a},
which is inspired by the Cherry-Hamiltonian
describing two counter-rotating harmonic oscillators \cite{Che1928}.
As exemplary system to study the transition from EE to CU stability
we use a variant of such two coupled
counter-rotating \twoD\ standard maps given by the map $\M(p_1, p_2, q_1,
q_2)\mapsto(p'_1, p'_2, q'_1, q'_2)$ as
\begin{subequations}
 \begin{align}
  p_1' &= p_1 + \frac{K_1}{2\pi}\sin{2\pi(q_1')} + \frac{K}{2\pi}
	  \sin{2\pi(q_1' + q_2')}\label{eq:mapping_a}\\
  p_2' &= p_2 + \frac{K_2}{2\pi}\sin{2\pi(q_2')} + \frac{K}{2\pi}
	  \sin{2\pi(q_1' + q_2')}\label{eq:mapping_b}\\
  q_1' &= q_1 + p_1\label{eq:mapping_c}\\
  q_2' &= q_2 - p_2,\label{eq:mapping_d}
 \end{align}
 \label{eq:mapping}%
\end{subequations}
where $K_1$ and $K_2$ are the kicking strengths of the two \twoD\ subsystems
and $K$ determines the coupling between them.
The phase space is restricted on the torus, i.e.\
$(p_1, p_2, q_1, q_2) \in [-0.5, 0.5[^2\times[0,1[^2$ with periodic boundary
conditions.
Note that the counter-rotating
character of the two uncoupled \twoD{} subsystems in $(p_1, q_1)$
and $(p_2, q_2)$ is due in the negative sign of the second momentum
$p_2$ in \eqnref{eq:mapping_d} which ensures
that condition \eqref{eq:Krein-DP-product} is fulfilled.
This sign is the only difference to the usual
\fourD{} standard map, as introduced in Refs.~\cite{Fro1970a, Fro1972}.
This map has also been investigated in Ref.~\cite{BaeMei2020},
though with the negative sign in \eqnref{eq:mapping_c}
instead of \eqnref{eq:mapping_d}.

\begin{figure}
 \includegraphics{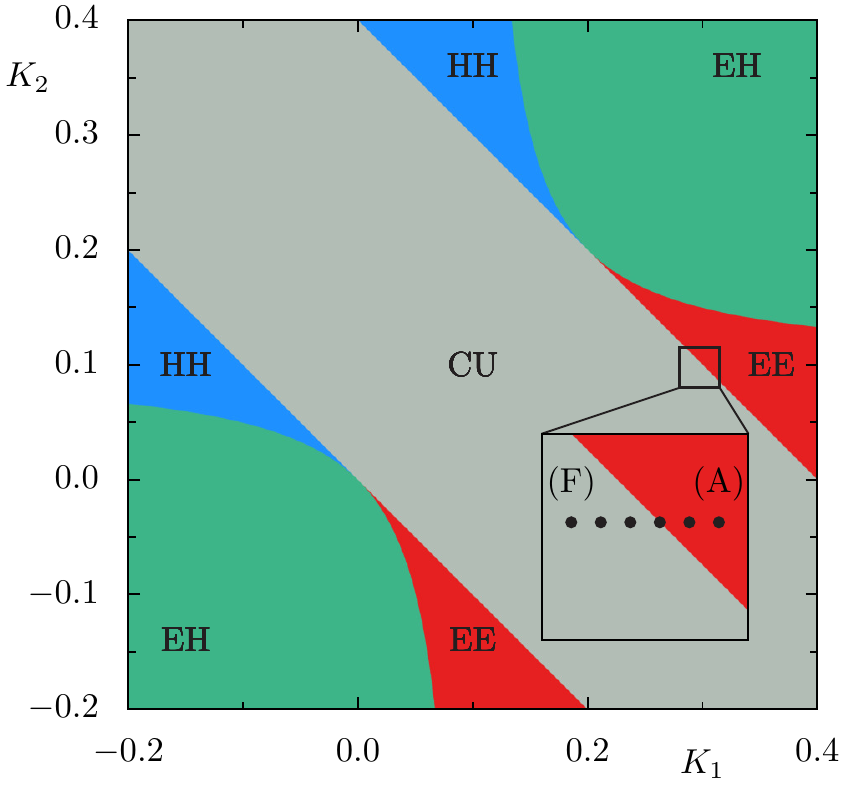}
 \caption{Stability of the fixed point
          $(0, 0, \nicefrac{1}{2}, \nicefrac{1}{2})$
          for fixed $K=0.1$ in dependence of $K_1$ and $K_2$.
          The magnification shows the selected parameters for
          the transition from EE to CU,
          (A) $K_1= 0.31$, (B) $K_1=0.305$, ..., and (F) $K_1=0.285$.
        }
 \label{fig:parameter}
\end{figure}

We will focus on the central fixed point at $\mv{z}^*=(0, 0, \nicefrac{1}{2},
\nicefrac{1}{2})$ in the following.
Its stability coefficients are
\begin{subequations}
 \begin{align}
  A &= -K_1 + K_2 + 4,\\
  B &= -K_1 K_2 + K_1 K - 2 K_1 + K_2 K + 2 K_2 + 6.
 \end{align}
 \label{eq:ab_coeffs}%
\end{subequations}
\Figref{fig:parameter} shows the stability diagram for fixed coupling $K=0.1$
in dependence on $K_1$ and $K_2$.
The fixed point is complex unstable in the region between the
two straight lines
\begin{equation}
 K_2=-K_1\quad\text{and}\quad K_2=4K-K_1.
\end{equation}
The saddle-center and the period-doubling boundaries,
Eq.~\eqref{eq:linear_boundaries}, lead to the hyperbolae
\begin{equation}
 -\frac{KK_1}{K-K_1}\quad\text{and}\quad\frac{-KK_1+4K_1-16}{K-K_1+4}.
\end{equation}
In order to investigate the transition from EE to CU stability, we choose the
EE region with positive kicking parameters and fix $K_2=0.1$
while $K_1$ is varied.
The six equidistant parameters $K_1=0.31, 0.305, ..., 0.285$,
are indicated as black points,
labeled by \pa\ to \pf\  with \pc\ lying directly on
Krein's boundary, Eq.~\eqref{eq:krein_parabola},
for $(K, K_1, K_2)=(0.1, 0.3, 0.1)$.

\begin{figure}
 \includegraphics{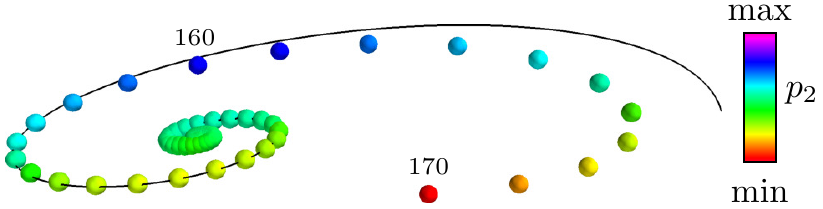}
 \caption{Spiraling motion of an orbit started close to the CU fixed point.
Shown are the $(p_1, q_1, q_2)$ coordinates with $p_2$ encoded in color of
the first 170 iterates of the point $(0, 0, 0.5, 0.5) + \mu$ for
$\mu=10^{-8}$. The initial spiraling motion
is well described using the linearized dynamics,
\eqnref{eq:eigenvector_expansion}, as shown by the black curve.
From the 160th iterate deviations become visible in the plot.
}
 \label{fig:spiral_motion}
\end{figure}

Once the fixed point has become complex unstable, we get a quadruplet of four
complex eigenvalues
$(\lambda, \lambda^{-1}, \bar{\lambda}, \bar{\lambda}^{-1})$ of $\D\M$
where $\lambda=\exp{(\beta+\mathrm{i}\theta)}$ with
$\beta\in\R_+$ and $\theta\in[0, \pi[$, see
Sec.~\ref{subsec:linear_stability}.
The corresponding eigenvectors
$(\xi_1, \xi_2, \bar{\xi}_1, \bar{\xi}_2)$
can be written as
$\xi_j=u_j+\mathrm{i}v_j$ with $u_j, v_j\in\R^4$ and $j=1,2$. The stable and
unstable invariant subspaces of the linearized map
are spanned by $u_1, v_1$ and $u_2, v_2$, respectively.

From this one key feature of the dynamics
in the surrounding of a CU fixed point follows:
Under the linearized dynamics
these eigenvectors evolve as $\xi_j^{n} = \lambda_j^{n}\xi_j$ and consequently
provides the evolution in the stable and unstable subspaces
by \cite{PapConPol1995}
\begin{subequations}
\begin{align}
 u_j^{(n)} &= \exp{\left(\pm\beta n\right)}\left(\cos{(n\theta)} u_j -
\sin{(n\theta)}v_j\right)\\
 v_j^{(n)} &= \exp{\left(\pm\beta n\right)}\left(\sin{(n\theta)} u_j +
\cos{(n\theta)}v_j\right),
\end{align}
\label{eq:eigenvector_expansion}%
\end{subequations}
where the positive sign corresponds to $j=1$ and the negative to $j=2$.
Any point $z$ in the \fourD{} phase space can be expressed
in the basis of the eigenvectors, i.e.\
$z=c_1 u_1 + c_2 v_1 + c_3 u_2 + c_4 v_2$
with coefficients $c_1, c_2, c_3, c_4\in\R$.
These coefficients can be determined with
the help of the basis of the dual space of the matrix of eigenvectors
\cite{RicLanBaeKet2014}.
Using the time evolution of the eigenvectors \eqnref{eq:eigenvector_expansion}
allows for obtaining the linearized dynamics of
an orbit for a given initial condition.
Apparently, the underlying dynamics is governed by an expanding/contracting
part and a rotating part which leads to a spiraling motion
as illustrated in \figref{fig:spiral_motion}.
If $c_1$ or $c_2$ are different from zero, the expanding dynamics
will asymptotically dominate.
Note that this provides a good description for some limited
number of iterations of the map $\M$ only, beyond which the nonlinear dynamics
becomes relevant, as can be seen by the
deviations between the real orbit depicted as colored spheres and the
linerized dynamics shown as black curve in \figref{fig:spiral_motion}.

\subsection{\threeDD\ phase-space slice}
\label{subsec:phase_space_slices}

To get an intuition for the dynamics of the transition from EE to CU
stability of the fixed point in phase space, we use a \threeD\
phase-space slice \cite{RicLanBaeKet2014}.
The idea is to reduce
the \fourD\ phase space by one dimension by considering a \threeD\ hyperplane
$\Gamma$ and determining those points of an orbit
that fulfill the slice condition
\begin{equation}
 \Gamma_\varepsilon = \big\{ (p_1, p_2, q_1, q_2) \; \big| \; |p_2| \le
\varepsilon \big\}.\label{eq:slice_condition}
\end{equation}
For the resulting points the coordinates $(p_1, q_1, q_2)$
are displayed in a \threeD{} plot.
The parameter $\varepsilon$, i.e., the thickness of the slice, controls the
resolution. Smaller values of $\varepsilon$ require
longer orbits to obtain the same number of points in the slice
as the slice condition \eqref{eq:slice_condition} is fulfilled less often.
For all \threeD\ phase-space slice plots in this paper
we choose $\varepsilon = 10^{-6}$.
Typically $f$-dimensional objects in the full \fourD{} phase space
appear as $(f-1)$-dimensional objects in the \threeD\ phase-space slice.
For example \twoD{} tori lead to two (or more) separate
(but dynamically connected) rings in the  \threeD\ phase-space slice
and \onetori{} lead to two (or more) points in the slice.
For further examples, also including more general slice conditions,
and detailed discussions see
Refs.~\cite{RicLanBaeKet2014,LanRicOnkBaeKet2014,OnkLanKetBae2016,
  LanBaeKet2016,AnaBouBae2017,FirLanKetBae2018,BaeMei2020}.

\begin{figure*}
 \centering
 \includegraphics{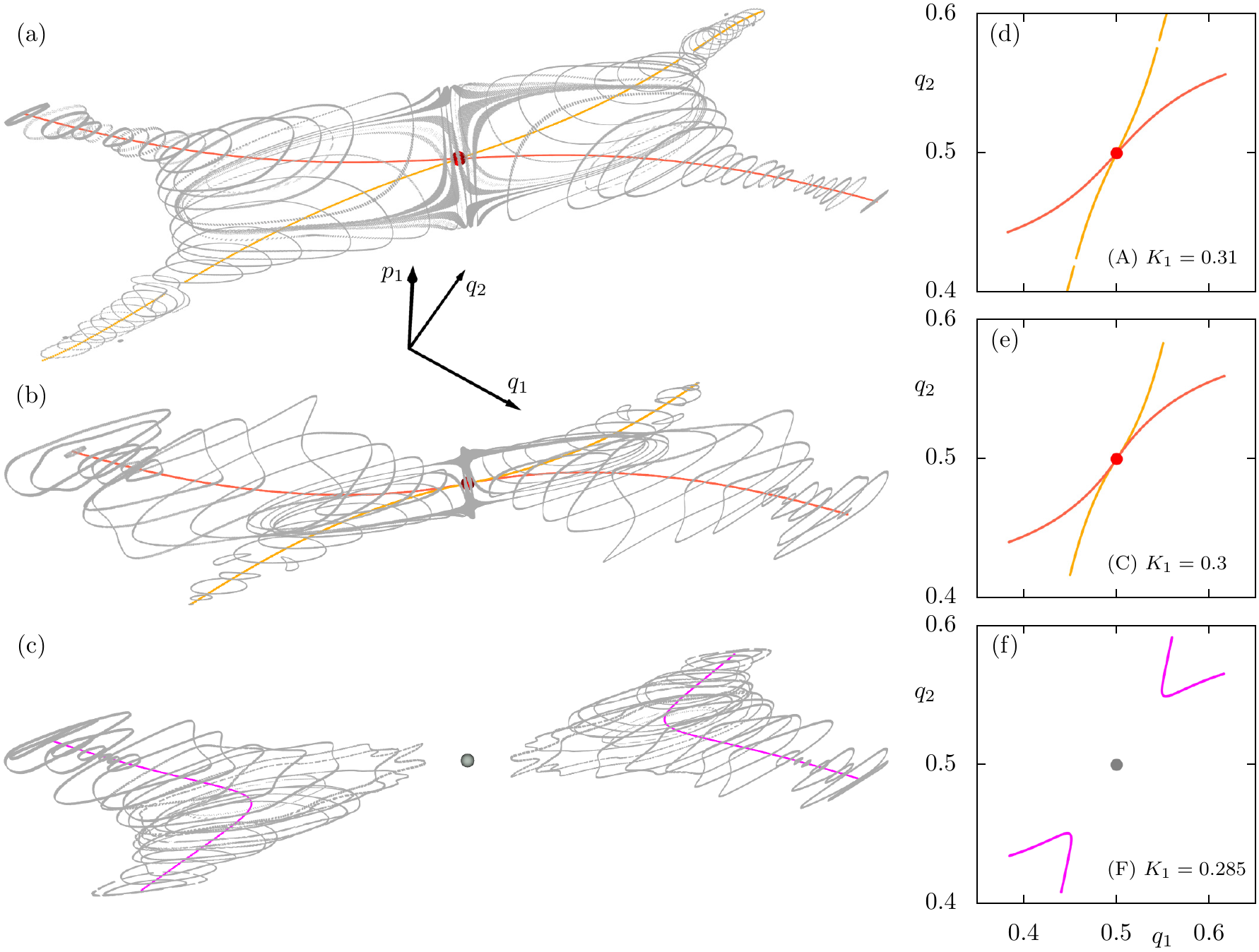}
 \caption{
   Sequence of \threeD\ phase-space slice plots of regular tori represented
as grey rings in the vicinity of the fixed point shown as red spheres for
elliptic-elliptic stability and as a grey sphere for complex instability.
The families of \oneD{} tori (red, yellow, magenta)
form the skeleton of the surrounding \twoD{} tori.
The chosen parameters are (a) $K_1=0.31$, (b) $K_1=0.3$, (c) $K_1=0.285$ and
correspond to points \pa, \pc\ and \pf\ in parameter space, see
\figref{fig:parameter}. The right column (d), (e), (f) depicts the families
of \onetori{}, which lie in the
$q_1$--$q_2$ plane due to the symmetry of the map.
\movierefall
}
 \label{fig:3d_slices}
\end{figure*}

\Figref{fig:3d_slices} shows a sequence of
\threeD\ phase-space slice plots of
regular orbits in the vicinity of the central fixed point for the parameter
sets \pa, \pc, and \pf, see \figref{fig:parameter}.
In \figref{fig:3d_slices}(a) for parameter set \pa, i.e.\
$K_1=0.31$, one is in the stable regime
and quite far away from the Krein collision.
The EE fixed point (red sphere)
is surrounded by regular \twoD{} tori shown as grey curves,
which form pairs closed loops on either side of the fixed point.
The general arrangement of the \twoD{}
tori is governed by the two (Lyapunov) families of \oneD{}-tori
which are attached to the EE fixed point
and shown in yellow and orange, respectively.
Due to the symmetries of the map, both families lie in the $q_1$--$q_2$ plane.
Thus they can be displayed in \twoD{} diagrams to clarify the change of
the families under parameter variation, see Figs.~\ref{fig:3d_slices}(d)-(f).
Note that the small gap in the yellow family in Fig.~\ref{fig:3d_slices}(a)
is caused by a resonance, see Sec.~\ref{subsec:frequencies}.
Both families of elliptic \oneD{}-tori are surrounded by regular 2-tori which
form pairs of rings in the \threeD\ phase-space representation, depicted
in gray color. Interestingly, the regular 2-tori in the direct vicinity of the
fixed point show a strong bending close to the fixed point.
This geometry
is similar to the phase space of the normal-form
for $\alpha>0$ in \figref{fig:poisson_map}(a) where the hyperbolic shape of the
$I\neq0$ plane forces the tori to bend away from the $y_1$--$y_2$ plane.
Furthermore, the families of \onetori{} correspond to the family of period-two
periodic points in the normal-form.

\Figref{fig:3d_slices}(b) shows the situation for point \pc\ in parameter
space with $K_1=0.3$. For this parameter the
two eigenvalue pairs of the linearized map at the fixed point
coalesce at two places on the complex unit circle,
see Fig.~\ref{fig:krein_collision}(b).
When approaching the Krein collision parameter, the angle between the
eigenvectors of the linearized map decreases until the eigenvectors of the
eigenvalue pairs become collinear.
Accordingly, the families of \onetori{}
are approximately parallel in the vicinity of the fixed point
as can be seen in \figref{fig:3d_slices}(b).

Finally, \figref{fig:3d_slices}(c) shows the
situation after the Krein collision, i.e.\
for $K_1=0.285$, which corresponds to parameter \pf\ in \figref{fig:parameter}.
Once the fixed point has become
complex unstable, the two families of \onetori{}
detach from the fixed point and merge into one single family.
This corresponds to the normal-form behavior for $\alpha<0$, see
\figref{fig:poisson_map}(b).
The regular tori close to the family of \onetori{} persist.
Interestingly, orbits in the vicinity of the CU fixed point stay in its
surrounding for very long times and only eventually escape. This will be
discussed in more detail in Sec.~\ref{sec:escape}.

\begin{figure}
 \includegraphics{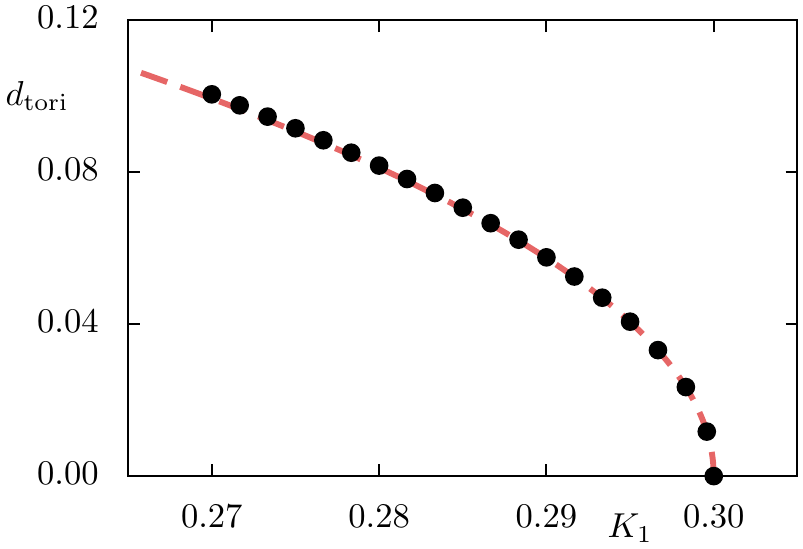}
 \caption{Minimal distance $d_{\text{tori}}$ between the central fixed
point and the family of \onetori{} in dependence on $K_1$. The distance follows
the predicted behavior $\propto\sqrt{K_1^*-K_1}$, shown as red dashed
line for $K_1\le K_1^*=0.3$.
}
 \label{fig:tori_distances}
\end{figure}

To quantify the detachment of the regular \twoD{} tori
from the CU fixed point, we compute the minimal distance \dtori\
between the complex unstable fixed point and the family of \onetori{}.
In the normal-form description of Sec.~\ref{subsec:normal_form}
the minimal distance is given by the distance between the trivial
fixed point at the origin and the period-two periodic point,
namely by $\dtori=\sqrt{-\alpha}/b$. For
the \fourD\ map this translates in first approximation to
\begin{equation} \label{eq:dtori-sqrt}
 \dtori\propto\sqrt{K_1^*-K_1}
\end{equation}
with $K_1 \le K_1^*=0.3$. \Figref{fig:tori_distances} shows the
numerically determined minimal distance \dtori\ in dependence on the kicking
strength $K_1$ as black dots. Good agreement with the square root behavior
\eqref{eq:dtori-sqrt}, shown as a dashed line, is found.
Further away from the Krein collision parameter small deviations
become visible.

\subsection{Frequency space}
\label{subsec:frequencies}

Complementary to the representation in phase space one can display regular
tori in frequency space, which is particularly useful for understanding
the influence of resonances.
A regular torus is characterized by two frequencies,
one describing the motion along the major
radius of the 2-torus and one for the motion along
the minor radius. Numerically the frequencies $\nu_1, \nu_2 \in [0, 1[$
for an orbit started in a phase-space point $(p_1(0), p_2(0), q_1(0), q_2(0))$
are determined using a Fourier-transform based frequency analysis
\cite{MarDavEzr1987,Las1990,Las1993,BarBazGioScaTod1996}.
As signals $z_j(n) = q_j(n) - \mathrm{i}p_j(n)$ for each degree
of freedom $j=1,2$ is used,  where $(q_j(n), p_j(n))$
are the coordinates obtained from $N$ successive iterates of the map.
In order to distinguish
regular and chaotic motion, the frequencies $\nu_j$ of the first half of
an orbit, i.e., iterates in the interval $n\in[0,N/2-1]$, are computed and
compared to the frequencies $\widetilde{\nu}_j$ of the second
half, i.e., the iterates in the interval $n\in[N/2,N]$.
For the motion on a regular torus, the
difference of these frequency pairs should be rather small.
Thus if the maximal difference
$\max\{|\nu_j - \widetilde{\nu}_j|\}$ is smaller than
some threshold $\delta_{\text{cut}}$, we consider the orbit as regular.
In the following $\delta_{\text{cut}}=10^{-8}$ is used.
Of course, such a numerical criterion does not guarantee
that the orbit eventually could become chaotic
at very large times, as is also the case with other chaos
indicators, see Ref.~\cite{SkoGotLas2016} for a recent overview.
Using an ensemble of $10^7$ initial conditions, randomly chosen in the
\fourD{} phase-space volume defined by $p_1, p_2\in[-0.1, 0.1]$ and $q_1,
q_2\in[0.4,0.6]$, and plotting the frequencies $(\nu_1, \nu_2)$
of the regular tori provides the two-dimensional
frequency space representation.

\begin{figure*}
 \includegraphics{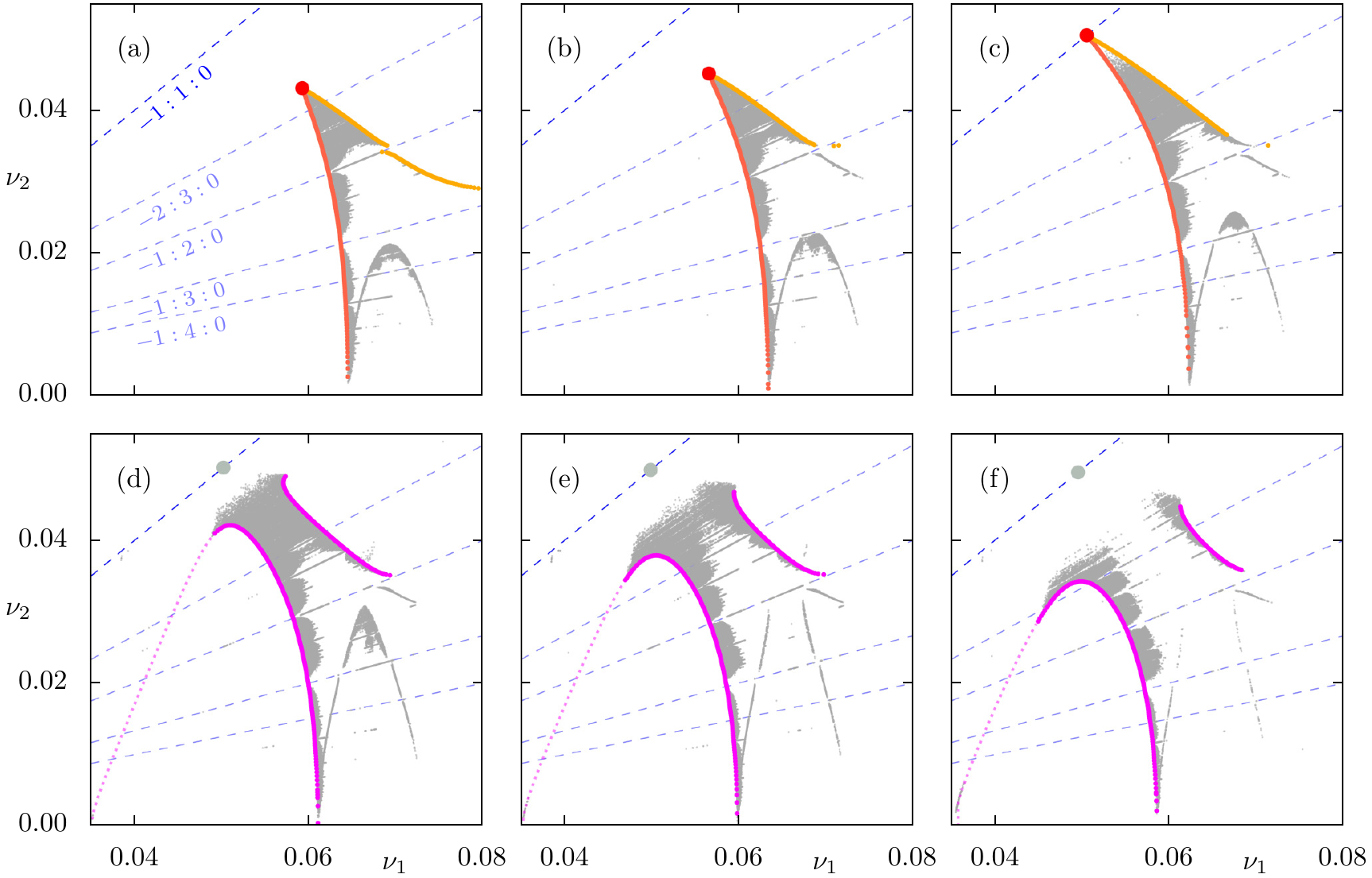}
 \caption{Frequency space for different parameters $K_1$:
    (a) $K_1= 0.31$, (b) $K_1=0.305$, ..., and (f) $K_1=0.285$,
    corresponding to (A)--(F) in \figref{fig:parameter}.
    Light gray dots correspond to regular orbits while
  the orange, yellow, and magenta dots correspond to the families of \oneD\
  tori. The frequency of the elliptic-elliptic (red) and complex
  unstable (gray) fixed point are depicted as enlarged dot.
  The dotted magenta curve in (d)-(f) is the unimodular transformation
  of the upper branch of \onetori{}. Some relevant resonance lines
  are shown as dashed lines.
}
 \label{fig:frequencies}
\end{figure*}

\Figref{fig:frequencies} shows a sequence of such frequency space plots for
all six parameter sets specified in \figref{fig:parameter}.
The frequencies of the EE fixed point is indicated
by a large red point in \figref{fig:frequencies}(a)-(c).
For the complex unstable fixed point there is only one frequency
given by the angle of the complex eigenvalues,
which is shown on the $-1:1:0$ resonance line
as large grey point in \figref{fig:frequencies}(d)-(f).
Although hardly noticable, the angle gets smaller with decreasing $K_1$.
As for the \threeD\ phase-space slice shown in \figref{fig:3d_slices},
the orange, yellow and magenta points mark the frequencies
of the families of \onetori{}, which form the edges of the gray regions
of regular tori.

Resonances correspond to straight lines in frequency space,
\begin{equation}
 n_1\nu_1+n_2\nu_2=m,
\end{equation}
with $m, n_1, n_2\in\Z$ and $\text{gcd}(m, n_1, n_2)=1$
and either $n_1 \neq 0$ or $n_2 \neq 0$.
Some relevant resonance lines are shown as blue dashed lines, labeled by
$n_1:n_2:m$. Such resonances lead to resonance channels \cite{Las1993} and
gaps in the families of \onetori{} \cite{LanRicOnkBaeKet2014}.

The typical frequency space around an EE fixed point
is seen in Fig.~\ref{fig:frequencies}(a)-(b)
for $K_1=0.31$ and $K_1=0.305$ which corresponds
to parameters \pa\ and \pb\ in \figref{fig:parameter},
respectively. Both families of \onetori{} are attached to the fixed point
forming a
cusp and the regular tori fill a region in between these families. As the
eigenvalues approach the Krein collision parameter in
\figref{fig:frequencies}(b),
the fixed point has to approach the $-1:1:0$ resonance line since the
eigenvalues of the linearized map
eventually coalesce on the pair $\mathrm{e}^{\pm\mathrm{i}2\pi\nu}$
with $\nu=\nu_1=\nu_2$.
This shift of the frequencies of the fixed point stretches the families
of \onetori{} and the top of the cusp accordingly. During this process, the
density
of regular tori close to the $-1:1:0$ resonance line decreases. This
becomes especially apparent in case of the Krein collision parameter in
\figref{fig:frequencies}(c), i.e.\ for parameter \pc\ in
\figref{fig:parameter} for $K_1=0.3$.
This corresponds to the tangency of the families of \onetori{}
so that only few regular tori exist in the surrounding
of the fixed point.

\Figref{fig:frequencies}(d)-(f) show the frequency space plots for the complex
unstable case for $K_1=0.295$, $K_1=0.29$, and $K_1=0.285$, corresponding
to the points \pd, \pe\ and \pf\ in \figref{fig:parameter}. The two former
families of \onetori{} merge in the Krein collision parameter and subsequently
detach
from the $-1:1:0$ resonance line once the fixed point looses its
stability. We observe two branches of the merged family which bend away from
the fixed point and simultaneously from the resonance line.
Note, that these branches are actually connected which can be seen
by applying the unimodular transformation
$(\nu_1, \nu_2)\mapsto(\nu_2, 2\nu_2-\nu_1)$ to the
upper branch resulting in the magenta dotted line. The transformed branch
connects seamlessly to the other branch yielding a complete arc beginning and
ending at $\nu_2\approx0$. This illustrates that both branches
actually belong to just one family of \onetori{}
after the fixed point has turned CU. In general, such linear transformations
with determinant $\pm1$ can always be applied for systems of periodic
functions \cite[Theorem 5 and 6]{Bor1927}.

Shortly after the transition of the fixed point to complex instability,
there are no regular tori
in its vicinity or the $-1:1:0$ resonance line in frequency space. However,
the regular tori between the branches of the former cusp still exist directly
after the transition as is visible in \figref{fig:frequencies}(d). Only when
the fixed point becomes more unstable, the distance of the branches
increases and the density of regular tori between them decreases until a gap
emerges, see \figref{fig:frequencies}(f).
The remaining regular orbits in \figref{fig:frequencies}(f)
are close to the family of \onetori{}. This
confirms the observations in the \threeD\ phase-space slice in
\figref{fig:3d_slices}(c), where regular tori are only found in the
surrounding of the family of \onetori{} and no regular structures are left in
the direct vicinity of the fixed point.

Note that the arc like structure in the range of $0.06\le\nu_1\le0.075$ below
the discussed region of regular tori, see Fig.~\ref{fig:frequencies}(a)-(d),
belongs to regular orbits in the surrounding of a periodic orbit close to the
central fixed point. Although these orbits are not in the focus of this study
they illustrate how the complex instability of the fixed point gradually
destroys all stable structures in its vicinity.

\section{Escape from the CU region}
\label{sec:escape}

\begin{figure*}
 \includegraphics{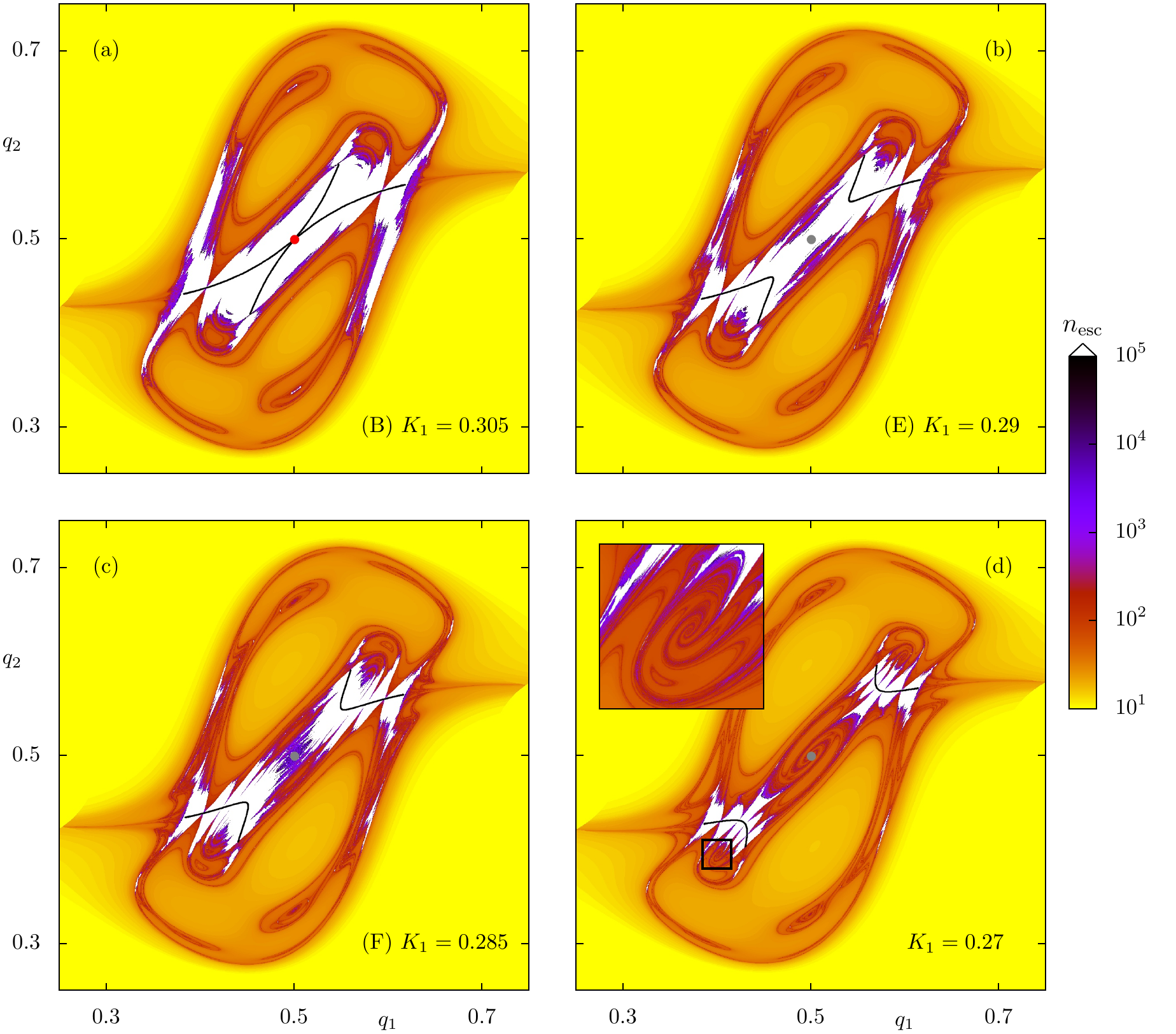}
 \caption{Escape time plots in the $q_1$--$q_2$ plane for $p_1=0$ and $p_2=0$
   for (a) $K_1=0.305$, (b) $K_1=0.29$, (c) $K_1=0.285$,
   and (d) $K=0.27$. The escape time is encoded in color,
   where white corresponds to those points which have not
   escaped within  $n_{\text{max}}=10^5$ iterations.
   The fixed point is shown as
   red (elliptic-elliptic) or gray dot (complex unstable)
   and the families of \onetori{} are shown as black dots.}
 \label{fig:escape}
\end{figure*}

When the EE fixed point becomes CU, this immediately affects its direct
surrounding as the two elliptic families of \onetori{}
become detached from the fixed point.
Thus there are also no regular tori in its direct vicinity.
Instead one has a \twoD\ stable and a \twoD\ unstable manifold which
lead to chaotic dynamics.
However, in practice close to the Krein collision parameter,
initial conditions in the vicinity of the fixed point
lead to orbits staying for very long times in a confined phase-space volume.
In this section, we investigate this behavior
and the underlying escape paths in more detail.

\subsection{Escape times}
\label{subsec:escape_times}

To study the escape of orbits from the surrounding of the CU fixed point, we
use escape time plots as in Refs.~\cite{GasRic1989, Sch2017, BaeMei2020},
see \figref{fig:escape}. Using a grid of
initial conditions on a particular plane in the \fourD\ phase space
for each initial point the escape time $n_{\text{esc}}$, required
to reach some specific exit region, is determined.
Since we are interested in the behaviour close to the
family of \onetori{}, we choose the initial points in the $q_1$--$q_2$ plane
through the fixed point with $q_1, q_2 \in [0.25,0.75]$ and $p_1=0$ and
$p_2=0$. On this plane, a $2000\times2000$ grid of initial points is used. We
define the exit region as
$q_1, q_2 \notin [0.25,0.75]$ with arbitrary momenta
$p_1, p_2\in[-0.5, 0.5]$.
Each initial condition
is iterated until it enters the exit region or a maximal number $n_{\text{max}}$
of iterations is reached.

\Figref{fig:escape} shows the escape time $n_{\text{esc}}$
encoded in color ranging from yellow for fast escaping
points to black for nearly regular orbits while white points do not escape to
the exit region within $n_{\text{max}}=10^5$ iterations
(though they may escape eventually).
In addition, the families of
\onetori{} are shown in black and the fixed point as red or gray dot for EE or
CU
stability, respectively.
The parameters for \figref{fig:escape}(a)-(c)
correspond to the points \pb, \pe,
and \pf\ in parameter space specified in \figref{fig:parameter}.
In addition, the families of
\onetori{} are shown in black and the fixed point as red or gray dot for EE or
CU
stability, respectively.

As before, we focus on the structures close to the fixed point. For the EE
case, the vicinity of the fixed point is naturally governed by a white region
which corresponds to the regular 2-tori surrounding the families of \onetori{},
compare with \figref{fig:escape}(a). Thus, even for arbitrarily large times
these orbits do not escape. Furthermore, we see the impact of the $-1:3:0$
resonance in form of a notch in the white region. This is consistent with the
frequency analysis in \figref{fig:frequencies}(a) for $K_1=0.31$.

If the eigenvalues of the fixed point approach the Krein collision
parameter the fraction of the white points only slightly diminishes and the
overall pattern of the escape time plot does not change much (not shown).
After the transition to complex
instability, see Figs.~\ref{fig:escape}(b)-(d), the white region reduces
substantially. Starting with the appearance of two small unstable regions in
the white region for $K_1=0.29$ in \figref{fig:escape}(b) above and below the
fixed point. Still, there are orbits in the direct vicinity of the fixed
point which stay close to it for more than $n_{\text{max}}$ iterations.
Quantitatively, the size of the white region depends on the threshold
$n_{\text{max}}$, but a larger value of $n_{\text{max}}$ does not affect the
shown escape time plots significantly.
The reason for this is that
orbits in the vicinity of the CU fixed point are confined for an extremely long
time when the parameters of the map are sufficiently close to the EE region in
\figref{fig:parameter}.
The more unstable the fixed point becomes, i.e.\ the
smaller $K_1$ is, the more the two branches of the family of \onetori{}
separate
and the white region diminishes because the regions of instability get larger.
Finally, for point \pf\ in \figref{fig:parameter} with $K_1=0.285$ all orbits
in the direct vicinity of the fixed point are able to reach the exit region
within $n_{\text{max}}$ iterations, see \figref{fig:escape}(c). For this
parameter we observe that the unstable regions in the escape plots reach the
fixed point, and consequently the large white region is divided into two smaller
ones. These two white regions correspond to the tubes of regular motion in the
\threeD\ phase-space slice representation, e.g. see \figref{fig:3d_slices}(f),
as well as the attached regular tori of the branches of the family of
\onetori{} in
frequency space, see \figref{fig:frequencies}(f).

\Figref{fig:escape}(d) shows the escape time plot for $K_1=0.27$, i.e.\ far
in the CU regime. The branches of the family of \onetori{} moved far away from
the
fixed point and the unstable region in between is large. Interestingly, this
unstable region reveals a unique spiral pattern which is attached to the fixed
point. Orbits on this spiral need at least one to two orders of magnitude more
iterations to escape into the exit region than the neighboring ones.
Additionally, there is another spiral structure on a smaller scale as shown
in the magnification in the inset.

A closer investigation of orbits started in the darker colored region reveals
that the spiral pattern is due to the influence of the $-2:3:0$
resonance: In cases where a frequency analysis of these orbits is
possible, i.e.\ the orbit is confined for long times and considered as regular
by our algorithm, see \secref{subsec:frequencies}, we get frequencies on
or close to this resonance line.

The escape time plots raise the following question:
Which structures govern the
slow transport in the vicinity of a complex unstable fixed point?
An important ingredient to answer this question are the invariant manifolds
of the fixed point, which are discussed in the next section.

\subsection{Stable and unstable manifolds}
\label{subsec:manifolds}

\begin{figure*}
\centering
 \includegraphics{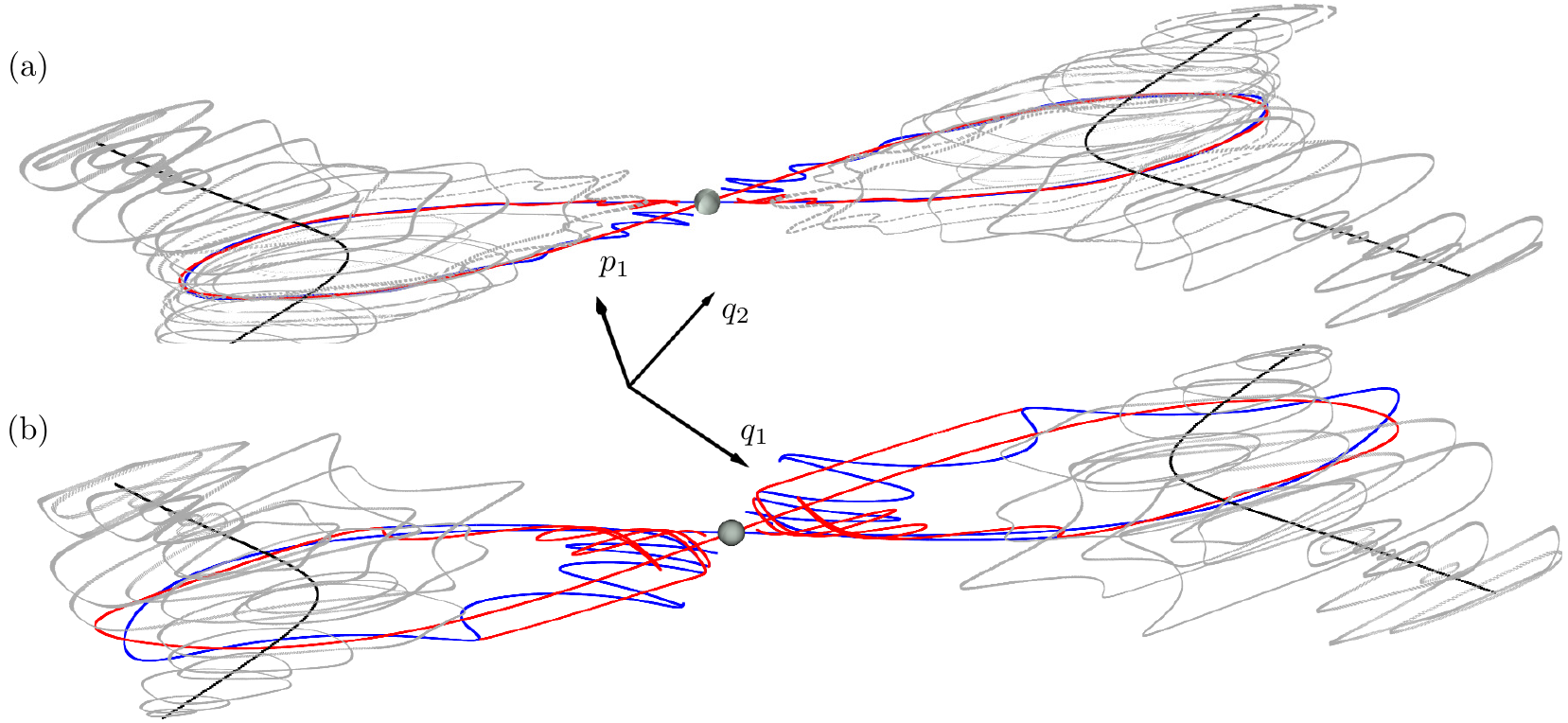}
 \caption{\threeD\ phase-space slice representation
   of the stable (blue) and unstable (red) manifolds of the CU fixed point
   together with regular 2-tori (gray) and the family of \onetori{} (black)
   for (a) $K_1=0.285$ and (b) $K_1=0.28$.
   Thus (a) corresponds to the
   point \pf\ in \figref{fig:parameter}, compare with
   \figref{fig:3d_slices}(c), (f).
   \movierefall}
 \label{fig:manifolds}
\end{figure*}

The stable and unstable manifolds associated with
an unstable fixed point govern the chaotic dynamics
in its surrounding.
For a complex unstable fixed point of a \fourD{} map
the manifolds are two-dimensional invariant objects
in the \fourD{} phase space.
Numerically the manifolds are computed using
the parameterization method \cite{CabFonLla2003a,
  CabFonLla2003b, CabFonLla2005,HarCanFigLuqMon2016, AnaBouBae2017, GonMir2017},
see Appendix~\ref{sec:manifold-computation} for details.
In the \threeD\ phase-space slice representation
they lead to one-dimensional curves, see \figref{fig:manifolds},
where the red curve corresponds to the unstable manifold
and the blue curve to the stable manifold.

The regular 2-tori (gray loops) as well as the families of
\onetori{} (black curves) in \figref{fig:manifolds}(a) are the same as in
\figref{fig:3d_slices}(f).
Figure~\ref{fig:manifolds}(b) shows the geometry
for a smaller value of $K_1=0.28$. The complex unstable fixed point is
indicated by a gray sphere in both plots.

Numerically it is found that the
stable and unstable manifolds intersect in one point.
This point therefore is a homoclinic point whose forward iterates
approach the fixed point on the stable manifold
while the backward iterates approach the fixed point on the unstable manifold.
The existence of a transverse homoclinic point therefore immediately
implies an infinity of such homoclinic points.
Note that generically two \twoD{} manifolds in a \fourD{} phase space will not
intersect. The fact that this happens for the manifolds of the
considered fixed point must be due to the symmetries of the map.

The geometry
becomes more clearly visible
for smaller  $K_1=0.28$ as shown in  \Figref{fig:manifolds}(b).
The arrangement of the manifolds in the \threeD{} phase-space slice
reminds of the homoclinic tangle in \twoD\ symplectic maps.
In comparison to \figref{fig:manifolds}(a)
the excursions of the manifolds are more pronounced
which corresponds to a larger chaotic region
surrounding the complex unstable fixed point.

It has to be emphasized, that even though the geometry
visually resembles the homoclinic tangle in \twoD{} symplectic maps,
the iterate of any of the homoclinic intersections
in general is not contained in the \threeD{} phase-space slice. Actually, we
find numerically that the stable and unstable manifolds intersect in a \oneD{}
line which is itself an invariant set.
Therefore, the intersection point in
the \threeD{} phase-space slice and its iterates are only a subset of the
\oneD{} intersection line.
Moreover, as the manifolds are only \twoD{} they cannot
enclose a volume, so that there is no equivalent
to the lobe structure and transport via a turnstile mechanism
as in \twoD{} symplectic maps \cite{KayMeiPer1984b,RomWig1990,Mei1992,Mei2015}.

\subsection{Escape statistics}
\label{subsec:escape_statistics}

\begin{figure}
 \includegraphics{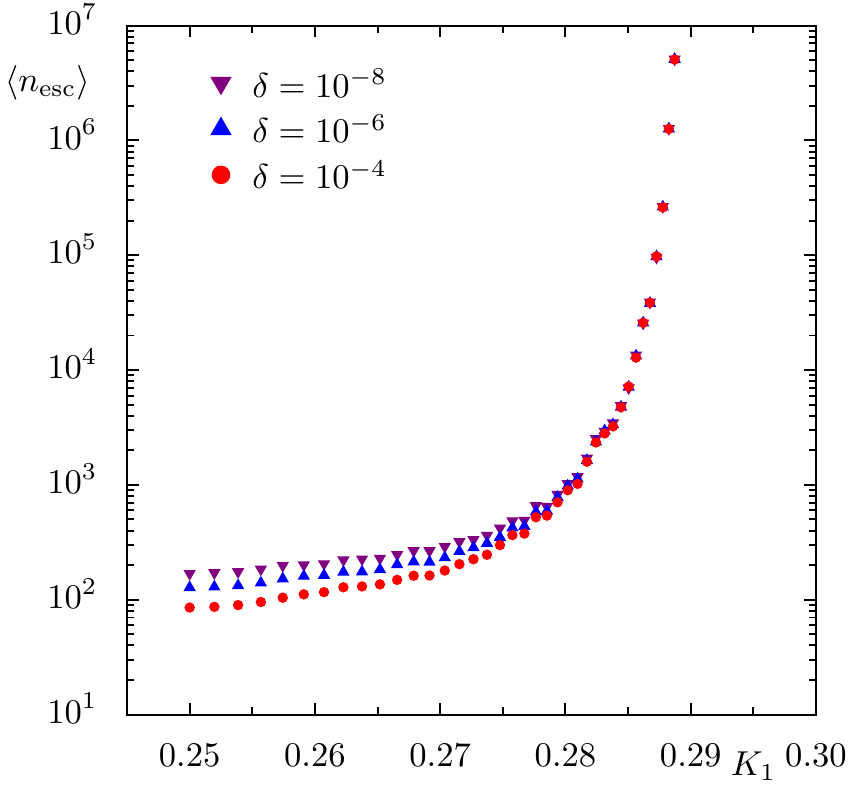}
 \caption{Average escape time $\langle n_{\text{esc}}\rangle$ of an
   ensemble of $10^4$ orbits started in $U_\delta$ in dependence on $K_1$
   for
   $\delta=10^{-8}$ (purple downward triangles),
   $\delta=10^{-6}$ (blue triangles), and
   $\delta=10^{-4}$ (red circles).}
 \label{fig:ensemble_escape_time}
\end{figure}

To investigate the chaotic transport in the vicinity of the CU
fixed point we consider an ensemble of initial conditions
in a \fourD\ cube
\begin{equation}
 U_\delta = [-\delta, \delta]^2 \times [0.5-\delta, 0.5+\delta]^2,
\end{equation}
with small $\delta$.
The exit region is again
chosen to be $p_1, p_2 \in [-0.5, 0.5]$ and $q_1, q_2 \notin [0.25, 0.75]$.
\Figref{fig:ensemble_escape_time} shows the average escape time
$\langle n_{\text{esc}}\rangle$ for an ensemble of $10^4$ orbits
in dependence on $K_1$
for different $\delta=10^{-4}$, $\delta=10^{-6}$, and $\delta=10^{-8}$.
When approaching the Krein collision parameter $K_1^*=0.3$,
the average escape time $\langle n_{\text{esc}}\rangle$
strongly increases and for $K_1 > 0.29$ exceeds $10^7$ iterations.
The same is also found for the smallest escape time (not shown).
Extracting the functional dependence from the data
turned out to inconclusive.

\begin{figure}
 \includegraphics{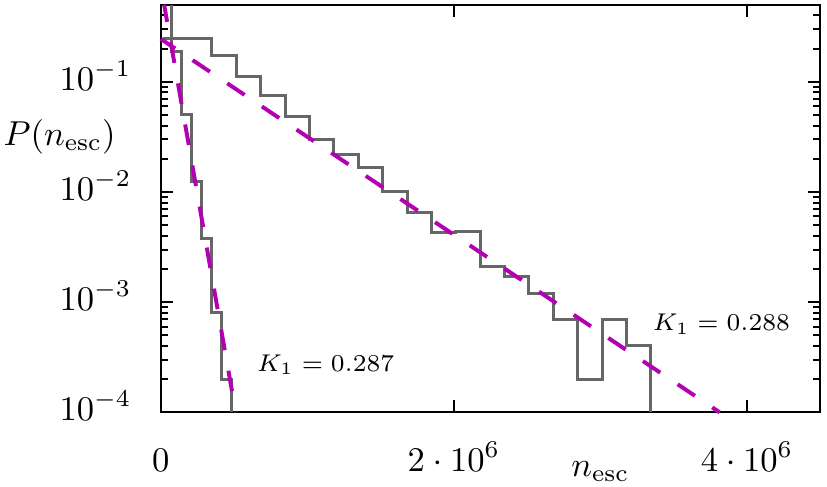}
 \caption{Histogram $P(n_{\text{esc}})$ of the escape times
   for $K_1 = 0.287$ and $K_1 = 0.288$.
   The dashed lines show a fit to an exponential for large $n_{\text{esc}}$
 }
 \label{fig:ensemble_escape_time_distribution}
\end{figure}

\begin{figure}
 \includegraphics{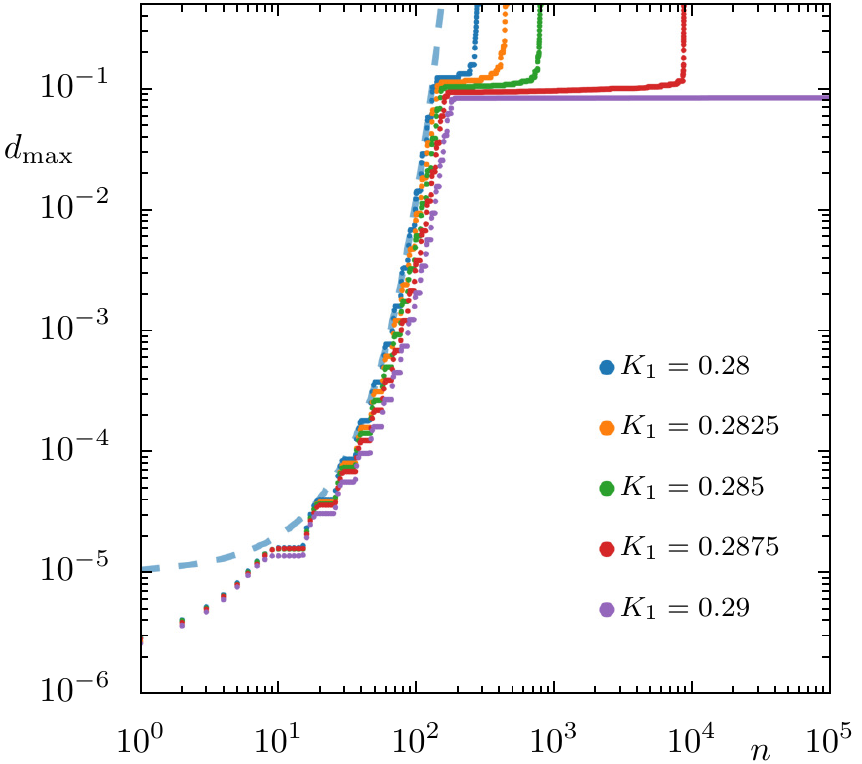}
 \caption{Maximal distance $d_{\text{max}}$ of an ensemble of $10^{4}$
initial conditions started in the \fourD{} cube $U_\delta$ with $\delta=10^{-6}$
vs. the number of iterations $n$
for $K_1=0.28, 0.2825, 0.285, 0.2875, 0.29$
(top to bottom, corresponding to increasing escape time).
The initial expansion is well described by $\propto|\lambda|^n$,
shown for $K_1=0.28$ (blue dashed curve).}
 \label{fig:ensemble_escape_distances}
\end{figure}

The tail of the distribution $P(n_{\text{esc}})$ of escape times is
very well described by an exponential,
see Fig.~\ref{fig:ensemble_escape_time_distribution}.
This provides a hint at what mechanism
could be responsible for such large escape times:
there could be one partial barrier (of unknown origin)
for the dynamics which allows for a small flux
towards the escape region \cite{Mei1992}.
Such a a single partial barrier would lead to a simple exponential
\cite{BauBer1990} while in contrast several partial barriers would typically
lead to an overall power-law
behavior \cite{DinBouOtt1990,ChiVec1993,She2010}.
Note that for the small hump of $\langle n_{\text{esc}}\rangle$ seen in
Fig.~\ref{fig:ensemble_escape_time} around $K_1=0.284$
the corresponding $P(n_{\text{esc}})$ shows a non-exponential
behavior in the tail.

To quantify the escape dynamics of the ensemble,
we now consider the extent as a function of the number of iterates.
Explicitly we determine
\begin{equation}
  d_{\text{max}}(n) = \text{max}_{i\le n} \{ ||\mv{z}^{(i)} - \mv{z}^* || \; | \;
  \text{ with } \mv{z}^{(0)} \in U_\delta \},
\end{equation}
where $\mv{z}^{(i)}$ is the $i$-th iterate of an
initial point $\mv{z}^{(0)}  \in U_\delta$
and $||\mv{z}_i  - \mv{z}^* ||$ is the distance to the
complex unstable fixed point at $\mv{z}^*$.
We use $10^4$ initial conditions in $U_\delta$ with $\delta=10^{-6}$.
\Figref{fig:ensemble_escape_distances} shows the result
for five different values of $K_1$.
The expansion during the first 100 iterations
is similar and after about 10 iterations follows an overall exponential
given by $|\lambda|^n$,  where $\lambda$ is the
eigenvalue with largest absolute value.
For $K_1=0.28$ this is illustrated by the blue dashed curve.
On a finer scale the initial expansion
happens in a step-like manner. This is due to the spiraling motion of each
orbit as illustrated in \figref{fig:spiral_motion}.
This motion has a different extent in the different directions,
so that a larger distance is only obtained
periodically after approximately ten iterations for the first
expansion phase.
This corresponds to half the reciprocal winding frequency of the fixed point.

After the first rapid expansion phase, the maximal distance
shows prominent plateaus extending over several orders of magnitude
in time.
These plateaus become longer the closer the parameter $K_1$ is
to $K_1^*=0.3$, i.e.\
the parameter of the Krein collision.
Thus for a very long time the ensemble is effectively
confined in phase space.
Afterwards there is at least one trajectory
which leaves this region very quickly,
as manifested by the sharp increase of $d_{\text{max}}$.

A closer look at the plateaus reveals that
there is still a rather slow increase.
The occurrence of the plateaus can be explained by
the alternating spiraling in and out of the dynamics already observed in
Refs.~\cite{Heg1985,JorOll2004, KatPatCon2011, DelCon2016}:
An orbit initially started near the complex unstable fixed point
moves away from it on a spiral
along the unstable manifold until it reaches a maximal distance to the fixed
point. This behavior corresponds to the first expansion phase up to
approximately 100 iterations.
Subsequently, the orbit spirals in again and
gets very close to the fixed point with some minimal distance.
When spiraling out again, this can lead
to a slightly increased maximal distance.
This process of inward
and outward spiraling repeats many times before the orbit escapes
quickly.
Note, that this sequence of outward and inward
spiraling only holds for parameters which are near the
elliptic-elliptic region in the parameter plot in \figref{fig:parameter},
i.e.\ if $K_1$ is sufficiently close to $K_1^*=0.3$.
Further away from the Krein collision parameter the extent
of the plateau of $d_{\text{max}}$ becomes very short or even non-existent,
see \figref{fig:ensemble_escape_distances} for $K_1=0.28$.

\begin{figure}
 \includegraphics{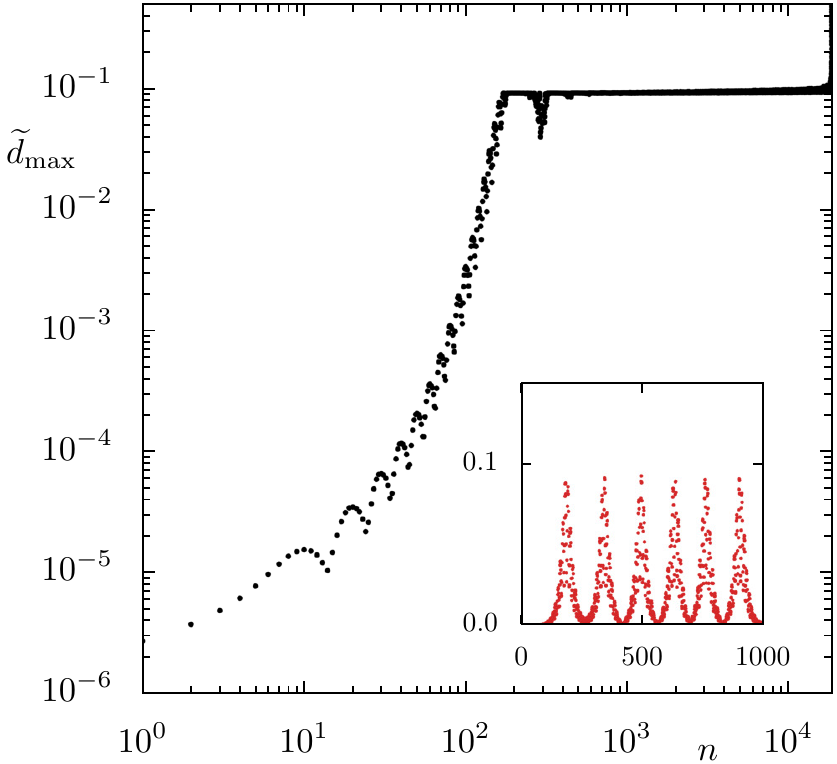}
 \caption{Maximal extent $\widetilde{d}_{\text{max}}$ of an ensemble of
$10^{4}$ initial conditions started in the \fourD{} cube $U_\delta$ with
$\delta=10^{-6}$ vs. the number of iterations $n$. The inset shows the maximal
extent of a single exemplary orbit up to the first 1000 iterations.}
 \label{fig:ensemble_extent}
\end{figure}

It is also illuminating to consider the extent of the iterated
ensemble at a given number of iterations,
\begin{equation}
  \widetilde{d}_{\text{max}}(n)
    = \text{max} \{ ||\mv{z}^{(n)} - \mv{z}^* || \; | \;
  \text{ with } \mv{z}^{(0)} \in U_\delta \},
\end{equation}
see Fig.~\ref{fig:ensemble_extent}.
Initially one has the overall exponential increase
which is superimposed by small oscillation caused by the spiraling
motion. This occurs until the ensemble has expanded until
the homoclinic intersection, which corresponds to the
beginning of the plateaus in \figref{fig:ensemble_escape_distances}.
Afterwards, there is a prominent dip around $n=200$
i.e.\ the extent of the ensemble has become quite small again
and most of the points are located in a small surrounding
of the complex unstable fixed point.
These minima
converge to the plateau for growing $n$ such that the second dip is already
barely visible. This effect is due to the inward and outward spiraling
behavior of each individual orbit.
The inset of \figref{fig:ensemble_extent}
shows $\widetilde{d}_{\text{max}}$ of one single orbit. The position of the
first minimum after the expansion of one single orbit matches
roughly the first minimum in the plateau of the ensemble.
This expansion and contraction of the ensemble repeats approximately
periodically until some loss of correlations sets in and
the dips of $d_{\text{max}}$ become  less and less prominent.
Note that such kind of dynamics is also found
for \twoD{} symplectic maps for the dynamics
after a period-doubling bifurcation and
also for \fourD{} symplectic maps with an II fixed-point.
A more detailed investigation and comparison of these cases
would be very interesting and is left for future studies.

\subsection{Escape dynamics}
\label{subsec:escape_dynamics}

\begin{figure}
 \centering
 \includegraphics{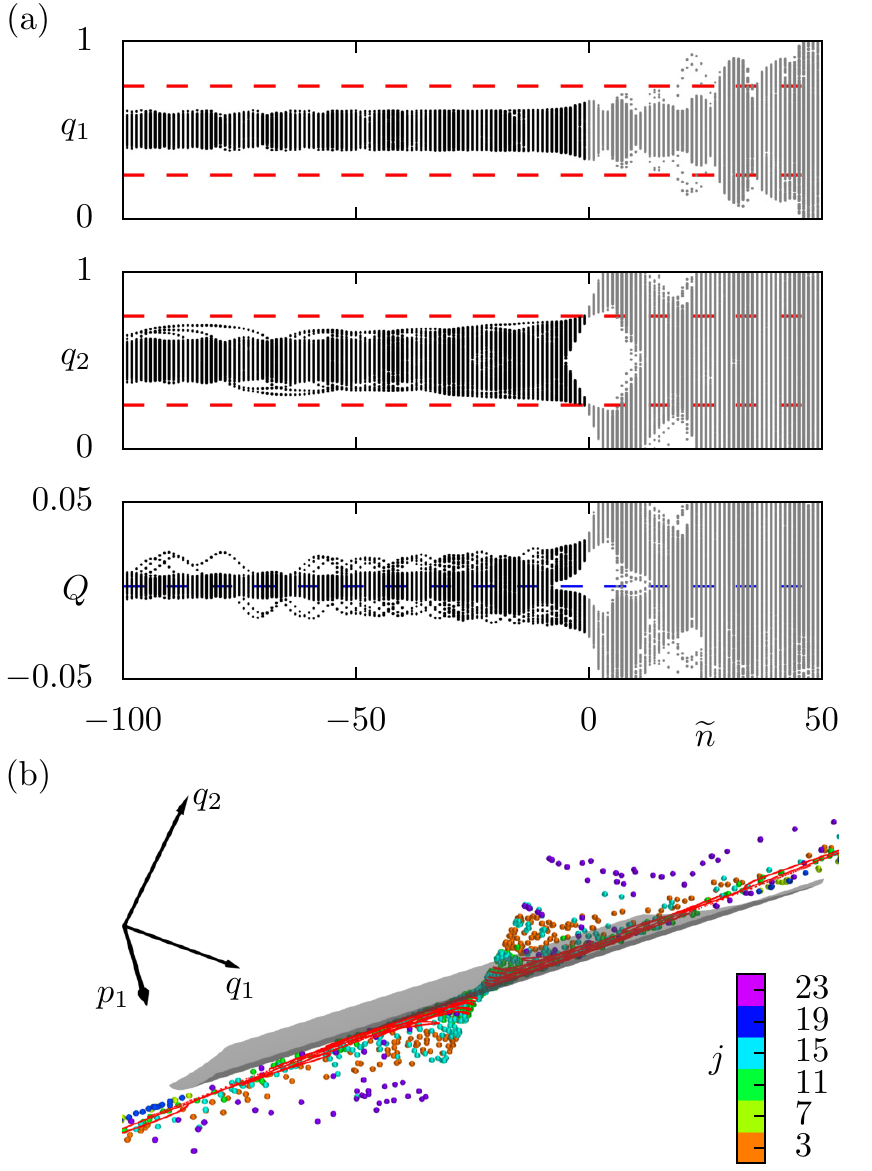}
 \caption{(a) Shown are the $q_1$, $q_2$-coordinates and the quadratic
 invariant of the linearization $Q$ of 1000 orbits started with random initial
 conditions in $U_\delta$ with $\delta=10^{-6}$ and $K_1=0.288$ over
 $\widetilde{n}=n-n_{\text{esc}}$. The escape criterion is the same as in the
 previous experiments and marked as a red dashed line while in the last plot
 the blue dashed line represents $Q_0$.
 (b) \threeD\ phase-space slice representation of segments of a single
 exemplary orbit for $K_1=0.288$.
 Each segment consists of consecutive 10000 iterates and
 shown are those points fulfilling the slice condition
 for the segments $j=3, 7, 11,
 15, 19$, and $23$, see the text for further explanation.
 The unstable manifold is
 shown as a red curve and the $Q=Q_0$ plane as gray transparent surface.
 \movierefall
 }
 \label{fig:orbit_drift}
\end{figure}

The temporal dependence of the extent of the iterates of the ensemble
allows for quantifying the long-time confinement
within the chaotic region surrounding the complex unstable fixed point.
Still, the key question is, what is responsible for this
long-time confinement and what is the escape mechanism?
In particular, referring to the normal form description,
there could be either an escape within the $I=0$ plane
or across different planes with $I\neq 0$.
Escape within $I=0$ would be
similar to the case of the period-doubling bifurcation in
\twoD{} symplectic maps, where just after the fixed point has become
unstable there are usually still invariant curves
so that an escape of orbits is only possible when
being further away from the bifurcation in parameter space.
In contrast, the escape across different planes with $I\neq 0$
would be a genuinely \fourD{} effect.
In principle there could also be a competition
between these two escape routes and which of them
is relevant could depend both on parameters and considered time-scales.

As a measure of the invariant $I$ of the normal form for a symplectic map
we make use of the quadratic invariant of the linearized map. With
\eqnref{eq:quadratic_invariant} we get
\begin{align}\label{eq:quadratic_invariant_explicit}
 Q = &-p_1^2 + p_2^2 -q_1^2 (K_1 - K) - q_2^2 (K_2-K)\nonumber\\
 &+ p_1q_1(K-K_1) + p_2q_2(K_2-K)\nonumber\\
 &+ K(p_1q_2 - p_2q_1 + 2q_1q_2).
\end{align}
By use of a suitable coordinate transformation
\eqnref{eq:quadratic_invariant_explicit} degenerates for the Krein collision
parameter into two planes, namely the $p_1=-p_2$ and the $q_1=q_2$ plane
\cite{Pfe1985a}.
These two planes geometrically correspond to the representation of the $I=0$
plane for the hyperplanes $x_2=0$ or alternatively $y_2=0$ in the normal form
description, see \secref{subsec:normal_form}. Hence, the
quadratic invariant at the fixed point is $Q(\bm{z}^*)=0$.

However, away from the Krein collision the two planes are not degenerate
anymore. Therefore $Q$ does not resemble the $I=0$ plane and we get
\begin{equation}
 Q_0 = Q(\bm{z}^*) = -\frac{K_1+K_2}{4} + K,
\end{equation}
which is not zero in general. Still it turns out, that
$Q-Q_0$ is a well suited quantity to approximate the invariant $I$ of
the normal form for a symplectic map.

To address the question of the possible escape route, it is helpful to
compare for an ensemble of initial conditions
the individual coordinates of the orbits right before they escape.
\Figref{fig:orbit_drift}(a) shows the $q_1$ and $q_2$ coordinates as well
as the quadratic invariant $Q$ as function of
$\widetilde{n} = n-n_{\text{esc}}$,
i.e.\ for a few iterations before and after the 
escape of an orbit.
The initial conditions of the ensemble with 1000 orbits are started in
$U_\delta$ with $\delta=10^{-6}$ and the kicking strength is $K_1=0.288$.
The orbits are confined for negative
$\widetilde{n}$ and fulfill the escape criterion for positive $\widetilde{n}$,
as indicated by the red dashed horizontal lines.
The spread of the distances of $q_1$ and $q_2$
around the fixed point, i.e. the width of the distribution of
distances around 0.5, is slightly increasing towards $\widetilde{n}=0$.
Even though this trend is visible in both coordinates,
the escape condition is reached first by the $q_2$ coordinate.

In order to understand the escape mechanism in terms of the phase space
geometry, we compare the escape path in the \threeD{} phase-space slice
with the geometry of the normal-form. The arrangement of regular
tori and the family of \onetori{}, see \figref{fig:3d_slices}, suggest that
the $Q=Q_0$ plane is a good approximation to the $I=0$ plane, compare to the
gray plane in \figref{fig:orbit_drift}(b).
Therefore, $Q-Q_0$
provides an approximate measure of how far a
point of an orbit is away from the $I=0$ plane,
see \figref{fig:orbit_drift}(a).
As for the single coordinates, $Q$ shows an overall increase
and is spread more widely as $\widetilde{n}$ approaches 0. However,
about 7 iterations before $\widetilde{n}=0$ the distribution of
$Q$ splits into two separate parts, away from 0.

\begin{figure}
\centering
 \includegraphics{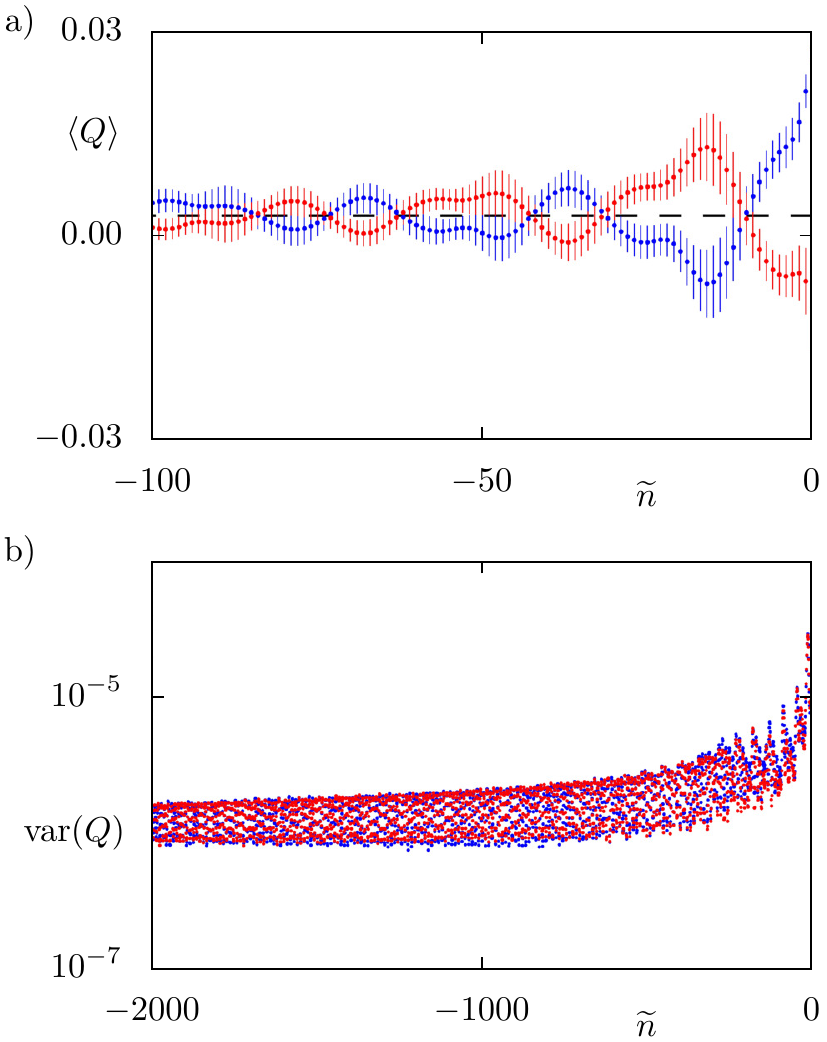}
 \caption{The mean width standard deviation a) and the variance b) of the
quadratic invariants $Q$ of the ensemble in \figref{fig:orbit_drift} are shown.
The blue set corresponds to the set of orbits with $Q(\widetilde{n}=0)>Q_0$ and
the inverse to the red data points. The black dashed line in a) represents
$Q_0$}
 \label{fig:quadratic_invariance_ensemble}
\end{figure}

In order to determine if the ensemble escapes through these two separated
escape paths or interchanges between those two, we split the ensemble in two
subsets by either $Q(\widetilde{n}=0)>Q_0$ or $Q(\widetilde{n}=0)<Q_0$ and
determine their mean and variance.
\Figref{fig:quadratic_invariance_ensemble}(a) shows the average as dots
and their standard deviation as error bars of the $Q(\widetilde{n}=0)>Q_0$
and the $Q(\widetilde{n}=0)<Q_0$ subset in blue and red color, respectively.
The ensemble clearly separates in these two sets and fluctuates around $Q_0$
marked as a black dashed line. Once the escape criterion is fulfilled, either
$Q>Q_0$ or $Q<Q_0$ and initially no further change in sign occurs. This
behavior translates to escape either across $I>0$ or $I<0$ planes in the
normal-form picture. Crossing the planes with different $I$
is only possible because the normal-form geometry
provided by Eq.~\eqref{eq:mapping} is broken.

\Figref{fig:quadratic_invariance_ensemble}(b) shows the time evolution of
the variance of both sets ranging from 2000 iterations before the escape up
to the escape. We observe the same type of
increase of the variance for both subsets
towards the escape at $\widetilde{n}=0$.
Understanding the behaviour of the variance quantitatively
is an interesting future task.

By following one single orbit we can also get an intuition of how the
orbit crosses the different $I\neq0$ planes, see \figref{fig:orbit_drift}(b).
Here, we consider a single orbit with initial condition
$(p_1, p_2, q_1, q_2) = (0,0,0.5+\mu,0.5+\mu)$ with $\mu=10^{-5}$
for $K_1=0.288$.
This orbit escapes after approximately 266000 iterations in our numerical
implementation of the map.
For this orbit we consider successive segments $[j \cdot 10000,
(j+1) \cdot 10000]$ of the iterates of the orbit.
For each segment those points fulfilling the
slice condition \eqref{eq:slice_condition} with $\varepsilon=10^{-4}$
are determined.
A selection in the surrounding of the complex unstable fixed point is shown in
\figref{fig:orbit_drift}(b) together with the \threeD{} phase-space slice of
the unstable manifold as a red curve.
This plot shows that the iterates of
the initial point are approximately restricted
around \oneD{} lines in the \threeD{} phase-space slice.
These lines follow the unstable manifold
and each of the successive segments appears to lie
on a slightly bent surface, similar to the $I\neq 0$ planes
of the normal form, compare with Fig.~\ref{fig:poisson_map}.
This suggests that an escaping orbit is following the unstable manifold
which gives rise to transport through the $I\neq0$ planes.
Note that the slice segments
for $j=7, 19$ are located at the excursion of the manifold farther away
from the fixed point and therefore do only appear at the edge of the
magnification.
In general, the motion along the unstable manifold explains also the
repetitive expanding and contracting behavior of the orbits.

\section{Summary and outlook}
\label{sec:summary}

In this paper the transition of a fixed point with elliptic-elliptic dynamics
to complex-unstable dynamics under parameter variation
is investigated for a \fourD{} symplectic map.
Using \threeD\ phase-space slices
we visualize regular dynamics in the vicinity of the fixed point. While
in the elliptic-elliptic case there exist two families of \onetori{} which are
attached to the fixed point and are surrounded by regular 2-tori, these
families merge into one single family and split off the fixed point.
Moreover, the geometry of regular orbits close to the fixed point
in the \threeD\ phase-space slice lie on surfaces
as predicted by the normal form description, see Fig.~\ref{fig:poisson_map}.
The phase-space representation is complemented by a frequency analysis of
regular tori, see Fig.~\ref{fig:frequencies}.
Before the transition to complex instability
the two families of \onetori{} are attached
to the fixed point forming a cusp-like region which encloses the regular tori.
The fixed point becomes complex unstable under parameter variation when
reaching the $-1:1:0$ resonance line
and the families of \onetori{} split off the fixed point.
Applying a unimodular transformation clarifies that these apparently two
families of \onetori{} actually form a single arc in frequency space.

Once the fixed point has become complex unstable
nearby orbits may eventually escape. However, it turns
out that shortly after the transition orbits are confined
to a particular phase-space region for very long times.
This region can be visualized using escape time plots,
see Fig.~\ref{fig:escape}.
The extent is governed by the stable and unstable invariant manifolds
of the complex unstable fixed point.
In the \threeD\ phase space they lead to a geometry which is visually similar
to that of the well-known homoclinic tangle for \twoD{} symplectic maps.

To quantify these observations we consider
the escape statistics for an ensemble of $10^4$ orbits,
started in the vicinity of the fixed point in dependence
on the distance to the bifurcation point, i.e. by varying the parameter $K_1$.
The average escape time strongly increases when approaching
the bifurcation point.
Measuring the maximal distance of all orbits of the ensemble
to the fixed point over the
number of iterations, reveals three different phases of the dynamics,
see Fig.~\ref{fig:ensemble_escape_distances}.
Initially, for the first approximately 100-200 iterations,
the distance increases exponentially, followed by a extended plateaus
in the second phase.
These plateaus correspond to the long-time confinement
and extend over longer times the closer the parameter is to the
Krein bifurcation.
A closer look at the plateaus shows that there
is a very slow increase as function of time.
The plateaus are due to the inward and outward spiraling dynamics
of the ensemble which follows the unstable invariant manifold. Thus, the
slope corresponds to a slowly growing extent of individual orbits,
see \figref{fig:ensemble_extent}.
Eventually, in the last phase one orbit of the ensemble will escape
after a critical time and the maximal distance of the ensemble
quickly reaches approximately 1.
If the fixed point is very unstable, the plateau is very short or even not
existent.

Comparing the $q_1$, $q_2$ coordinates and the quadratic invariant $Q$
of the ensemble for the transition from phase two to three
allows for determining the main escape paths close to the bifurcation,
see \figref{fig:orbit_drift}. This provides evidence
that long confined orbits escape across either $I>0$ or $I<0$ planes of the
normal-form. Thus the escape mechanism
is genuinely higher-dimensional.

Based on the improved understanding
of the geometry and escape of orbits near a complex unstable
fixed point, an interesting future task
is to explicitly determine the invariant $I$
for the specific map using a numerical normal form analysis.
This would allow for accurately quantifying
the transport across the approximately invariant planes
and to investigate whether the escape can be described
by a diffusive process.

\begin{acknowledgments}
  We are grateful for discussions
  with Markus Firmbach, Franziska H\"ubner, Roland Ketzmerick, and
  Haris Skokos.
  Robert MacKay kindly provided us with a copy of Ref.~\cite{BriCusMac1995}.
  Furthermore, we acknowledge support by the Deutsche Forschungsgemeinschaft
  under grant KE~537/6--1.

  All \threeDD{} visualizations were created using
  \textsc{Mayavi}~\cite{RamVar2011}.

\end{acknowledgments}

\appendix
\section{Computing stable and unstable manifolds}
\label{sec:manifold-computation}

There are various methods to determine the invariant manifolds
associated with an unstable fixed point,
see e.g.\ Refs.~\cite{YouKosYor1991, Hob1993, KosYorYou1996,GooWro2011,
WroGoo2013,EftConKat2014}.
Here we use the parametrization method, which was introduced in
Refs.~\cite{CabFonLla2003a, CabFonLla2003b, CabFonLla2005}
and has been used for example in
Refs.~\cite{HarCanFigLuqMon2016, AnaBouBae2017, GonMir2017}.

The parameterization method takes advantage of the
Hartman-Grobman theorem which for symplectic maps states that the
linearization of a fixed point or periodic orbit is conjugate to its
local stable and unstable invariant manifolds $W^{\text{s},
\text{u}}_{\text{loc}}$,
if the eigenvalues have an absolute value different from one, i.e.,
if they are unstable.
The key point of the parameterization method is to find
smooth vector-valued functions $\mathcal{F}_{\text{s}}$ and
$\mathcal{F}_{\text{u}}$ which parameterize the stable and
unstable invariant manifolds.
In order to do so, $\mathcal{F}_{\text{s},\text{u}}$ have to
obey on the one hand the linear conditions
\begin{align}
 \label{eq:lin_cond_1}
 \mathcal{F}_{s,u}(\bm{0}) &= \bm{z}^*\quad\text{and}\\[5pt]
 \label{eq:lin_cond_2}
 \frac{\partial\mathcal{F}_{s,u}(\bm{\theta})}{\partial\theta_j} &=
\bm{\xi}_j\qquad\text{for}\qquad 1\le i\le n_{s,u}
\end{align}
with $\bm{\theta}=(\theta_1, \;\ldots\;,
\theta_{n_{s,u}})\in\C^{n_{s,u}}$ and
$\xi_j\in\C^{2n_{s,u}}$ being the associated
eigenvector to the $n_s$ stable
and $n_u$ unstable eigenvalues $\lambda_j$.

On the other hand, $\mathcal{F}_{s,u}$ must satisfy the conjugacy equation
\begin{equation}
 \M\circ\mathcal{F}_{s,u}(\bm{\theta}) = \mathcal{F}_{s,u}(\lambda_1\theta_1,
\;\ldots\;,
 \lambda_{n_{s,u}}\theta_{n_{s,u}})
 \label{eq:conjugacy_equation}
\end{equation}
in order to take the non-linearity of the map into account.

For a complex unstable fixed point
of a \fourD{} symeplectic map one finds $n_s=n_u=2$.
Therefore, we expand $\mathcal{F}_{s,u}$ into the power series
\begin{equation}
 \mathcal{F}_{s,u}(\theta_1, \theta_2)=
 \begin{pmatrix}p_1(\theta_1, \theta_2)\\
  p_2(\theta_1, \theta_2)\\ q_1(\theta_1, \theta_2)\\ q_2(\theta_1, \theta_2)
  \end{pmatrix}=
 \sum\limits_{i=0}^{\infty}
 \sum\limits_{j=0}^{\infty}\bm{f}_{ij}\;\theta_1^i\;\theta_2^j,
\label{eq:parameterization_series_expansion}
\end{equation}
with vector-valued coefficients $\bm{f}_{ij}\in\C^4$.

For the considered map \eqref{eq:mapping},
the non-linear terms consist of sine functions with
various input arguments, namely three different
sums of phase-space coordinates. We approximate these sine
functions  by their Taylor series representation.
Advantageously, the coefficients
of this series can be easily computed by an
auto-differentiation algorithm
which is based on Refs.~\cite{Nei1992, Kje2014}.
Using the series representation of the sine terms of the
map $\M$ and combining \eqref{eq:parameterization_series_expansion}
and the conjugacy equation \eqref{eq:conjugacy_equation} leads to a
homological equation which can be solved iteratively for the coefficients
$\bm{f}_{ij}$ up to a given order $(m,n)$. The corresponding initial value 
problem is solved by the linear conditions \eqref{eq:lin_cond_1} and 
\eqref{eq:lin_cond_2}.

\FloatBarrier


\end{document}